\documentclass[useAMS,usenatbib]{mn2e}

\usepackage[T1]{fontenc}
\usepackage{ae,aecompl}
\usepackage[dvips]{graphicx}
\usepackage{epsfig}
\usepackage{rotating}
\usepackage{graphicx}
\usepackage[caption=false]{subfig}
\usepackage{color}
\usepackage{hyperref}

\newcommand{\hMsol}{\, h^{-1}{\rm M_{\odot}}}
\newcommand{\hMpc}{\, h^{-1}{\rm Mpc}}

\title[The accretion history of dark halos II]
{The accretion history of dark matter halos II:\\ The connections with the mass power spectrum and the density profile} 
\author[C.A.~Correa, J.S.B.~Wyithe, J.~Schaye and A.R.~Duffy]
  {Camila A.~Correa$^1$\thanks{E-mail: correac@student.unimelb.edu.au}, J. Stuart B.~Wyithe$^1$, Joop~Schaye$^2$ and Alan R.~Duffy$^{1,3}$\\
  $^1$ School of Physics, University of Melbourne, Parkville, VIC 3010, Australia\\
  $^2$ Leiden Observatory, Leiden University, PO Box 9513, 2300 RA Leiden, The Netherlands\\
  $^3$ Centre for Astrophysics and Supercomputing, Swinburne University of Technology, Melbourne, VIC 3122, Australia}
\date{\today}

\pagerange{\pageref{firstpage}--\pageref{lastpage}} \pubyear{2013}

\def\LaTeX{L\kern-.36em\raise.3ex\hbox{a}\kern-.15em
    T\kern-.1667em\lower.7ex\hbox{E}\kern-.125emX}

\begin{document}

\label{firstpage}

\maketitle

\begin{abstract}
We explore the relation between the structure and mass accretion histories of dark matter halos using a suite of cosmological simulations. We confirm that the formation time, defined as the time when the virial mass of the main progenitor equals the mass enclosed within the scale radius, correlates strongly with concentration. We provide a semi-analytic model for halo mass history that combines analytic relations with fits to simulations. This model has the functional form, $M(z)=M_{0}(1+z)^{\alpha}e^{\beta z}$, where the parameters $\alpha$ and $\beta$ are directly correlated with concentration. We then combine this model for the halo mass history with the analytic relations between $\alpha$, $\beta$ and the linear power spectrum derived by Correa et al. to establish the physical link between halo concentration and the initial density perturbation field. Finally, we provide fitting formulae for the halo mass history as well as numerical routines\footnote{}, we derive the accretion rate as a function of halo mass, and demonstrate how the halo mass history depends on cosmology and the adopted definition of halo mass.
\end{abstract}


\footnotetext{Available at \href{https://bitbucket.org/astroduff/commah}{\it{https://bitbucket.org/astroduff/commah}}} 

\section{Introduction}

Dark matter halos provide the potential wells inside which galaxies form. As a result, understanding their basic properties, including their formation history and internal structure, is an important step for understanding galaxy evolution. It is generally believed that the halo mass accretion history determines dark matter halo properties, such as their `universal' density profile (\citealt{NFW96}, hereafter NFW). The argument is as follows. During hierarchical growth, halos form through mergers with smaller structures and accretion from the intergalactic medium. Most mergers are minor (with smaller satellite halos) and do not alter the structure of the inner halo. Major mergers (mergers between halos of comparable mass) can bring material to the centre, but they are found not to play a pivotal role in modifying the internal mass distribution (\citealt{Wang}). Halo formation can therefore be described as an `inside out' process, where a strongly bound core collapses, followed by the gradual addition of material at the cosmological accretion rate. Through this process, halos acquire a nearly universal density profile that can be described by a simple formula known as the `NFW profile' (NFW). 

The origin of this universal density profile is not fully understood. One possibility is that the NFW profile results from a relaxation mechanism that produces equilibrium and is largely independent of the initial conditions and merger history (\citealt{NFW97}). However, another popular explanation, originally proposed by \citet{Syer}, is that the NFW profile is determined by the halo mass history, and it is then expected that halos should also follow a universal mass history profile (\citealt{Dekel03,Manrique,Sheth,Dalal,Salvador,Giocoli}). This universal accretion history was recently illustrated by \citet{Ludlow}, who showed that the halo mass histories, if scaled to certain values, follow the NFW profile. This was done by comparing the mass accretion history, expressed in terms of the critical density of the Universe, $M(\rho_{\rm{crit}}(z))$, with the NFW density profile, expressed in units of enclosed mass and mean density within $r$, $M(\langle\rho\rangle(<r))$ at $z=0$, in a mass-density plane.

In this work we aim to provide a model that links the halo mass history with the halo {\it{concentration}}, a parameter that fully describes the internal structure of dark matter halos. By doing so, we will gain insight into the origin of the NFW profile and its connection with the halo mass history. We also aim to find a physical explanation for the known correlation between the linear {\it{rms}} fluctuation of the density field, $\sigma$, and halo concentration.

This paper is organized as follows. We briefly introduce our simulations in Section \ref{SimulationData}, where we explain how we calculated the merger history trees and discuss the necessary numerical convergence conditions. Then we provide a model for the halo mass history, which we refer to as the {\it{semi-analytic model}}. This semi-analytic model is described in Section \ref{MAHmodel}, along with an analysis of the formation time definition. For this model we use the empirical \citet{McBride} formula. This functional form was motivated by EPS theory in a companion paper (\citealt{PaperI}; hereafter Paper I), and we calibrate the correlation between its two-parameters ($\alpha$ and $\beta$) using numerical simulations. As a result, the semi-analytic model combines analytic relations with fits to simulations, to relate halo structure to the mass accretion history. In Section \ref{Discussion} we show how the semi-analytic model for the halo mass history depends on cosmology and the adopted definition for halo mass. In Section \ref{compare_models} we provide a detailed comparison between the semi-analytic halo mass history model provided in this work, and the analytic model presented in Paper I. The parameters in this analytic model depend on the linear power spectrum and halo mass, whereas in the semi-analytic model the parameters depend on concentration and halo mass. We therefore combine the two models to establish the physical relation between the linear power spectrum and halo concentration. We will expand on this in a forthcoming paper (\citealt{PaperIII}, hereafter Paper III), where we predict the evolution of the concentration-mass relation and its dependence on cosmology. Finally, we provide a summary of formulae and discuss our main findings in Section \ref{SummaryAndConclussion}.

\section{Simulations}\label{SimulationData}

In this work we use the set of cosmological hydrodynamical simulations (the REF model) along with a set of dark matter only (DMONLY) simulations from the OWLS project (\citealt{Schaye}). These simulations were run with a significantly extended version of the N-Body Tree-PM, smoothed particle hydrodynamics (SPH) code Gadget3 (last described in \citealt{Springel05}). In order to assess the effects of the finite resolution and box size on our results, most simulations were run using the same physical model (DMONLY or REF) but different box sizes (ranging from $25\hMpc$ to $400\hMpc$) and particle numbers (ranging from $128^{3}$ to $512^{3}$). The main numerical parameters of the runs are listed in Table 1. The simulation names contain strings of the form LxxxNyyy, where xxx is the simulation box size in comoving $\hMpc$, and yyy is the cube root of the number of particles per species (dark matter or baryonic). For more details on the simulations we refer the reader to Appendix \ref{Sims_details}.

Our DMONLY simulations assume the WMAP5 cosmology, whereas the REF simulations assume WMAP3. To investigate the dependence on the adopted cosmological parameters, we include an extra set of five dark matter only simulations ($100\hMpc$ box size and $512^{3}$ dark matter particles) which assume values for the cosmological parameters derived from the different releases of Wilkinson Microwave Anisotropy Probe (WMAP) and the {\it{Planck}} mission. Table \ref{cosmo} lists the sets of cosmological parameters adopted in the different simulations. 

\begin{table*}
\centering  
\caption{List of simulations.  From
  left-to-right the columns show: simulation identifier; comoving box size;
  number of dark matter particles (there are equally 
  many baryonic particles); initial baryonic particle mass; dark matter
  particle mass; comoving (Plummer-equivalent) gravitational
  softening; maximum physical softening; final redshift. } 
\label{sims}
\begin{tabular}{lrrrlrrl}
\hline
  Simulation & L & N & $m_{\rm{b}}$ & $m_{\rm{dm}}$ & $\epsilon_{\rm{com}}$ & $\epsilon_{\rm{prop}}$ & $z_{\rm{end}}$ \\ 
   & ($h^{-1}\rm{Mpc}$) & & ($h^{-1}\rm{M}_{\sun}$) & ($h^{-1}\rm{M}_{\sun}$) & ($h^{-1}\rm{kpc}$) & ($h^{-1}\rm{kpc}$) &  \\  \hline\hline
   REF$_{-}$L100N512 & 100 & $512^{3}$ & $8.7\times 10^{7}$ & $4.1\times 10^{8}$ & 7.81 & 2.00 & $0$\\
   REF$_{-}$L100N256 & 100 & $256^{3}$ & $6.9\times 10^{8}$ & $3.2\times 10^{9}$ & 15.62 & 4.00 & $0$\\
   REF$_{-}$L100N128 & 100 & $128^{3}$ & $5.5\times 10^{9}$ & $2.6\times 10^{10}$ & 31.25 & 8.00 & $0$\\
   REF$_{-}$L050N512 & 50 & $512^{3}$ & $1.1\times 10^{7}$ & $5.1\times 10^{7}$ & 3.91 & 1.00 & $0$\\
   REF$_{-}$L025N512 & 25 & $512^{3}$ & $1.4\times 10^{6}$ & $6.3\times 10^{6}$ & 1.95 & 0.50 & $2$\\
   REF$_{-}$L025N256 & 25 & $256^{3}$ & $1.1\times 10^{7}$ & $5.1\times 10^{7}$ & 3.91 & 1.00 & $2$\\
   REF$_{-}$L025N128 & 25 & $128^{3}$ & $8.7\times 10^{7}$ & $4.1\times 10^{8}$ & 7.81 & 2.00 & $2$\\
   DMONL$Y_{-}$WMAP5$_{-}$L400N512 & 400 & $512^{3}$ & $-$ & $3.4\times 10^{10}$ & 31.25 & 8.00 & $0$\\
   DMONL$Y_{-}$WMAP5$_{-}$L200N512 & 200 & $512^{3}$ & $-$ & $3.2\times 10^{9}$ & 15.62 & 4.00 & $0$\\
   DMONL$Y_{-}$WMAP5$_{-}$L100N512 & 100 & $512^{3}$ & $-$ & $5.3\times 10^{8}$ & 7.81 & 2.00 & $0$\\
   DMONL$Y_{-}$WMAP5$_{-}$L050N512 & 50 & $512^{3}$ & $-$ & $6.1\times 10^{7}$ & 3.91 & 1.00 & $0$\\
   DMONL$Y_{-}$WMAP5$_{-}$L025N512 & 25 & $512^{3}$ & $-$ & $8.3\times 10^{6}$ & 2.00 & 0.50 & $0$\\ \hline
\end{tabular}
\end{table*}

\begin{figure*}
\begin{center}
\includegraphics[width=0.8\textwidth]{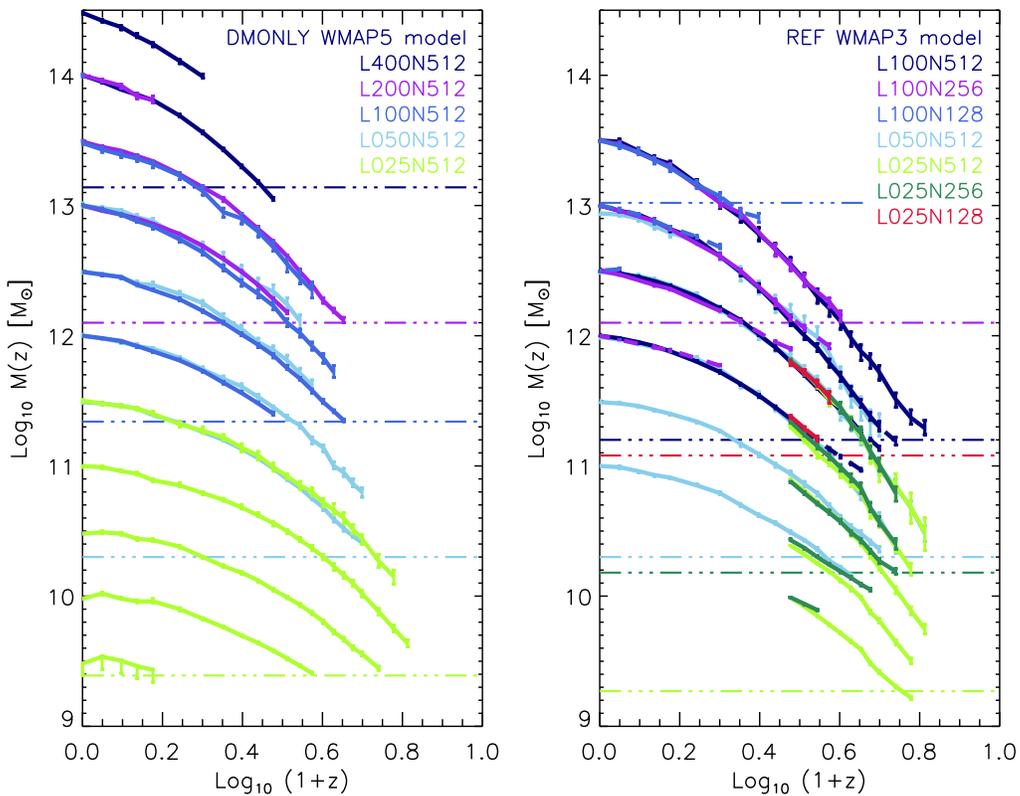}
\caption{Median halo mass history as a function of redshift from simulations DMONLY (left panel) and REF (right panel) for halos in eleven and seven different mass bins, respectively. The curves show the median value, and the $1\sigma$ error bars are determined by bootstrap resampling the halos from the merger tree at a given output redshift. The different colour lines show the mass histories of halos from different simulations. We find that a necessary condition for a halo to be defined, and mass histories to converge, is that halos should have a minimum of 300 dark matter particles. The horizontal dashed dotted lines show the $300\times m_{\rm{dm}}$ limit for the simulation that matches the colour, where $m_{\rm dm}$ is the respective dark matter particle mass. When following a merger tree from a given halo sample, some halos are discarded when unresolved. This introduces a bias and so an upturn in the median mass history. Therefore, mass history curves are stopped once fewer than $50\%$ of the original sample of halos are considered. Simulations in the REF model with $25\hMpc$ comoving box size have a final redshift of $z=2$, therefore the halos mass histories begin at this redshift.}
\label{fig1}
\end{center}
\end{figure*}

Halo mass histories are obtained from the simulation outputs by building halo merger trees. We define the halo mass history as the mass of the most massive halo (main progenitor) along the main branch of the merger tree. The method used to create the merger trees is described in detail in Appendix \ref{merger_tree}. While analysing the merger trees from the simulations, we look for a numerical resolution criterion under which mass accretion histories converge numerically. We begin by investigating the minimum number of particles halos must contain so that the merger trees lead to accurate numerical convergence. We find a necessary minimum limit of 300 dark matter particles, which corresponds to a minimum dark matter halo mass of $M_{\rm{halo}}\sim 2.3\times 10^{11}\rm{M}_{\sun}$ in the $100 \hMpc$ box, $M_{\rm{halo}}\sim 2.6\times 10^{10}\rm{M}_{\sun}$ in the $50\hMpc$ box, and $M_{\rm{halo}}\sim 3.4\times 10^{9}\rm{M}_{\sun}$ in the $25 \hMpc$ box.

In a merger tree, when a progenitor halo contains less than 300 dark matter particles, it is considered unresolved and discarded from the analysis. As a result, the number of halos in the sample that contribute to the median value of the mass history decreases with increasing redshift. Removing unresolved halos from the merger tree can introduce a bias. When the number of halos that are discarded drops to more than $50\%$ of the original sample, a spurious upturn in the median mass history occurs. To avoid this bias, the median mass history curve is only built out to the redshift at which less than $50\%$ of the original number of halos contribute to the median mass value. 

\begin{table}
\centering  
\caption{Cosmological parameters.} 
\label{cosmo}
\begin{tabular}{lllllll}
\hline
  Simulation & $\Omega_{\rm{m}}$ & $\Omega_{\Lambda}$ & $h$ & $\sigma_{8}$ & $n_{s}$ \\ \hline\hline
   DMONLY$_{-}$WMAP1 & 0.25 & 0.75 & 0.73 & 0.90 & 1.000\\
   DMONLY$_{-}$WMAP3 & 0.238 & 0.762 & 0.73 & 0.74 & 0.951\\
   DMONLY$_{-}$WMAP5 & 0.258 & 0.742 & 0.72 & 0.796 & 0.963\\
   DMONLY$_{-}$WMAP9 & 0.282 & 0.718 & 0.70 & 0.817 & 0.964\\
   DMONLY$_{-}${\it{Planck}}1 & 0.317 & 0.683 & 0.67 & 0.834 & 0.962\\

\hline
\end{tabular}
\end{table}

Fig. \ref{fig1} shows the effects of changing the resolution for the dark matter only and reference simulations. We first vary the box size while keeping the number of particles fixed (left panel). Then we vary the number of particles while keeping the box size fixed (right panel). The left panel (right panel) of Fig. \ref{fig1} shows the mass history as a function of redshift for halos in eleven (seven) different mass bins for the DMONLY (REF) simulation. All halo masses are binned in equally spaced logarithmic bins of size $\Delta \log_{10}M=0.5$. The mass histories are computed by calculating the median value of the halo masses from the merger tree at a given output redshift, the error bars correspond to 1$\sigma$ confidence intervals. The different coloured lines indicate the different simulations from which the halo mass histories were calculated. The horizontal dash-dotted lines in the panels show the $300\times m_{\rm{dm}}$ limit for the simulation that matches the colour. Halos in the simulation with masses lower than this value are unresolved, and hence their mass histories are not considered. The mass histories from halos whose main progenitors have masses lower than $10^{12}\rm{M}_{\sun}$ at $z=0$ were computed from simulations with $50\hMpc$ and $25\hMpc$ comoving box sizes. In the right panel, where the mass histories from the REF model are shown, all mass history curves obtained from the REF simulation with a $25\hMpc$ comoving box size have a final redshift of $z=2$. Therefore, these halo mass histories begin at this redshift.

\section{Semi-analytic model for the halo mass history}\label{MAHmodel}

In the following subsections we study dark matter halo properties and provide a semi-analytic model that relates halo structure to the mass accretion history. We begin with the NFW density profile and derive an analytic expression for the mean inner density, $\langle\rho\rangle(<r_{-2})$, within the scale radius, $r_{-2}$. We then define the formation redshift, and use the simulations to find the relation between $\langle\rho\rangle(<r_{-2})$ and the critical density of the universe at the formation redshift. We discuss the universality of the mass history curve and show how we can obtain an semi-analytic model for the mass history that depends on only one parameter (as expected from our EPS analysis presented in a companion paper). We then calibrate this single parameter fit using our numerical simulations. Finally, we show how the semi-analytic model for halo mass history depends on cosmology and halo mass definition.

\subsection{Density profile}

An important property of a population of halos is their spherically averaged density profile. Based on $N$-body simulations, \citet{NFW97} found that the density profiles of CDM halos can be approximated by a two parameter profile, 

\begin{equation}\label{NFW}
\rho(r)=\frac{\rho_{\rm{crit}}\delta_{c}}{(r/r_{-2})(1+r/r_{-2})^{2}},
\end{equation}
where $r$ is the radius, $r_{-2}$ is the characteristic radius at which the logarithmic density slope is $-2$, $\rho_{\rm{crit}}(z)=3H^{2}(z)/8\pi G$ is the critical density of the universe and $\delta_{c}$ is a dimensionless parameter related to the concentration $c$ by

\begin{equation}
\delta_{c}=\frac{200}{3}\frac{c^{3}}{[\ln(1+c)-c/(1+c)]},
\end{equation}

\noindent which applies at fixed virial mass and where $c$ is defined as $c=r_{200}/r_{-2}$, and $r_{200}$ is the virial radius. A halo is often defined so that the mean density $\langle\rho\rangle(<r)$ within the halo virial radius $r_{\Delta}$ is a factor $\Delta$ times the critical density of the universe at redshift $z$. Unfortunately, not all authors adopt the same definition, and readers should be aware of the difference in halo formation history and internal structure when different mass definitions are adopted (see \citealt{Duffy08,Diemer}). We explore this in Section \ref{Massdef} to which the reader is referred to for further details. Throughout this work we use $\Delta = 200$. We denote $M_{z}\equiv M_{200}(z)$ as the halo mass as a function of redshift, $M_{r}\equiv M(<r)$ as the halo mass profile within radius $r$ at $z=0$, $r_{200}$ as the virial radius at $z=0$ and $c$ as the concentration at $z=0$. Note that the halo mass is defined as all matter within the radius $r_{200}$ (see Table \ref{Notation} for reference).

The NFW profile is characterized by a logarithmic slope that steepens gradually from the centre outwards, and can be fully specified by the concentration parameter and halo mass. Simulations have shown that these two parameters are correlated, with the average concentration of a halo being a weakly decreasing function of mass (e.g. NFW; \citealt{Bullock,Eke,Shaw,Neto,Duffy08,Maccio08,Dutton14,Diemer14}). Therefore, the NFW density profile can be described by a single free parameter, the concentration, which can be related to virial mass. The following relation was found by \citet{Duffy08} from a large set of $N-$body simulations with the WMAP5 cosmology, 

\begin{equation}\label{c-m}
c = 6.67(M_{200}/2\times10^{12}h^{-1}\rm{M}_{\sun})^{-0.092},
\end{equation} 

\noindent for halos in equilibrium (relaxed).

\begin{table}
\centering  
\caption{Notation reference. Unless specified otherwise, quantities are evaluated at $z=0$.} 
\label{Notation}
\begin{tabular}{ll}
\hline
  Notation & Definition \\ \hline\hline
$M_{200}$ & $M_{r}(r_{200})$, total halo mass\\
$r_{200}$ & Virial radius\\
$r_{-2}$ & NFW scale radius\\
$c$ & NFW concentration\\
$M_{z}$ & $M(z)$, total halo mass at redshift $z$\\
$M_{r}(r)$ & $M(<r)$, mass enclosed within $r$\\
$x$ & $r/r_{200}$\\
$\langle\rho\rangle(<r_{-2})$ & Mean density within $r_{-2}$\\
$M_{r}(r_{-2})$ & $M(<r_{-2})$, enclosed mass within $r_{-2}$\\
$z_{-2}$ & Formation redshift, when $M_{z}$ equals $M_{r}(r_{-2})$\\
$\rho_{\rm{crit},0}$ & Critical density\\
$\rho_{\rm{crit}}(z)$ & Critical density at redshift $z$\\
$\rho_{\rm{m}}(z)$ & Mean background density at redshift $z$\\
\hline
\end{tabular}
\end{table}

The NFW profile can be expressed in terms of the mean internal density
\begin{equation}\label{rho_mean}
\langle\rho\rangle(<r) =\frac{M_{r}(r)}{(4\pi/3)r^{3}}=\frac{200}{x^{3}}\frac{Y(cx)}{Y(c)}\rho_{\rm{crit}},
\end{equation}
where $x$ is defined as $x=r/r_{200}$ and $Y(u)=\ln(1+u)-u/(1+u)$. From this last equation we can verify that at $r=r_{200}$, $x=1$ and $\langle\rho\rangle(<r_{200})=200\rho_{\rm{crit}}$.

Evaluating $\langle\rho\rangle(<r)$ at $r=r_{-2}$, we obtain
\begin{equation}\label{rho-2}
\langle\rho\rangle(<r_{-2})=\frac{M_{r}(r_{-2})}{(4\pi/3)r_{-2}^{3}}=200c^{3}\frac{Y(1)}{Y(c)}\rho_{\rm{crit}}.
\end{equation}

\noindent From this last expression we see that for a fixed redshift the mean inner density ${\langle\rho\rangle(<r_{-2})}$ can be written in terms of $c$. By substituting eq.~(\ref{c-m}) into (\ref{rho-2}), we can obtain ${\langle\rho\rangle(<r_{-2})}$ as a function of virial mass. Finally, we can compute the mass enclosed within $r_{-2}$. From eq.~(\ref{rho-2}) we obtain

\begin{equation}\label{M2}
M_{r}(r_{-2}) = M_{200}\frac{Y(1)}{Y(c)},
\end{equation}

\noindent where we used $M_{200}=(4\pi/3)r_{200}^{3}200\rho_{\rm{crit}}$.

Although the NFW profile is widely used and generally describes halo density profiles with high accuracy, it is worth noting that high resolution numerical simulations have shown that the spherically averaged density profiles of dark matter halos have small but systematic deviations from the NFW form (e.g. \citealt{Navarro04,Hayashi,Navarro10,Ludlow10,Diemer14a}). While there is no clear understanding of what breaks the structural similarity among halos, an alternative parametrization is sometimes used (the Einasto profile), which assumes the logarithmic slope to be a simple power law of radius, $d\ln\rho/d\ln r\propto (r/r_{-2})^{\alpha}$ (\citealt{Einasto}). Recently, \citet{Ludlow} investigated the relation between the accretion history and mass profile of cold dark matter halos. They found that halos whose mass profiles deviate from NFW and are better approximated by Einasto profiles also have accretion histories that deviate from the NFW shape in a similar way. However, they found the residuals from the systematic deviations from the NFW shape to be smaller than $10\%$. We therefore only consider the NFW halo density profile in this work.    

\begin{figure} 
\begin{center}
\includegraphics[angle=0,width=0.48\textwidth]{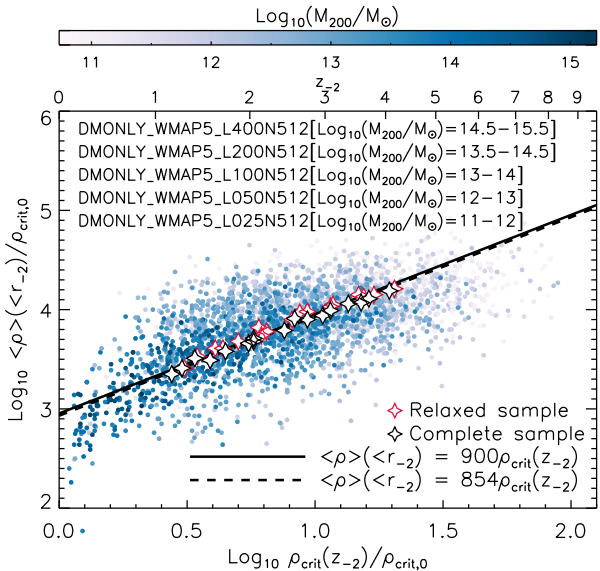}%
\caption{Relation between the mean density within the NFW scale radius at $z=0$ and the critical density of the universe at the halo formation redshift, $z_{-2}$, for DMONLY simulations from the OWLS project. The simulations assume the WMAP-5 cosmological parameters and have box sizes of 400$\hMpc$, 200$\hMpc$, 100$\hMpc$, 50$\hMpc$ and 25$\hMpc$. The black solid line indicates the relation shown in eq.~(\ref{rho-crit}), which only depends on cosmology through the mass-concentration relation. The black (red) star symbols show the mean values of the complete (relaxed) sample in logarithmic mass bins of width $\delta\log_{10}M=0.2$. The black dashed and solid lines show the relations found by fitting the data of the complete and relaxed samples, respectively. The filled circles correspond to values of individual halos and are coloured by mass according to the colour bar at the top of the plot.} 
\label{density_relation}
\end{center}
\end{figure}

\subsection{Formation redshift}

\citet{NFW97} showed that the characteristic overdensity ($\delta_{\rm{c}}$) is closely related to formation time ($z_{\rm{f}}$), which they defined as the time when half the mass of the main progenitor was first contained in progenitors larger than some fraction $f$ of the mass of the halo at $z=0$. They found that the `natural' relation $\delta_{\rm{c}}\propto \Omega_{\rm{m}}(1+z_{\rm{f}})^{3}$ describes how the overdensity of halos varies with their formation redshift. Subsequent investigations have used N-body simulations and empirical models to explore the relation between concentration and formation history in more detail (\citealt{Wechsler,Zhao03,Zhao09}). A good definition of formation time that relates concentration to halo mass history was found to be the time when the main progenitor switches from a period of fast growth to one of slow growth. This is based on the observation that halos that have experienced a recent major merger typically have relatively low concentrations, while halos that have experienced a longer phase of relatively quiescent growth have larger concentrations. Moreover, \citet{Zhao09} argue that halo concentration can be very well determined at the time the main progenitor of the halo has $4\%$ of its final mass.

The various formation time definitions each provide accurate fits to the simulations on which they are based and, at a given halo mass, show reasonably small scatter. However, our goal is to adopt a formation time definition that has a natural justification without invoking arbitrary mass fractions. To this end, we go back to the idea that halos are formed `inside out', and consider the formation time to be defined as the time when the initial bound core forms. We follow Ludlow et al. (2013) and define the formation redshift as the time at which the mass of the main progenitor equals the mass enclosed within the scale radius at $z=0$, yielding 

\begin{equation}
z_{-2}=z[M_{z}=M_{r}(r_{-2})].
\end{equation}

\noindent From now on we denote the formation redshift by $z_{-2}$. Interestingly, \citet{Ludlow} found that at this formation redshift, the critical density of the universe is directly proportional to the mean density within the scale radius of halos at $z=0: \rho_{\rm{crit}}(z_{-2})\propto \langle\rho\rangle(<r_{-2})$. A possible interpretation of this relation is that the central structure of a dark matter halo (contained within $r_{-2}$) is established through collapse and later accretion and mergers increase the mass and size of the halo without adding much material to its inner regions, thus increasing the halo virial radius while leaving the scale radius and its inner density ($\langle\rho\rangle(<r_{-2})$) almost unchanged (\citealt{Huss,Wang}).

\begin{figure*} 
  \centering
 \subfloat{\includegraphics[angle=0,width=0.48\textwidth]{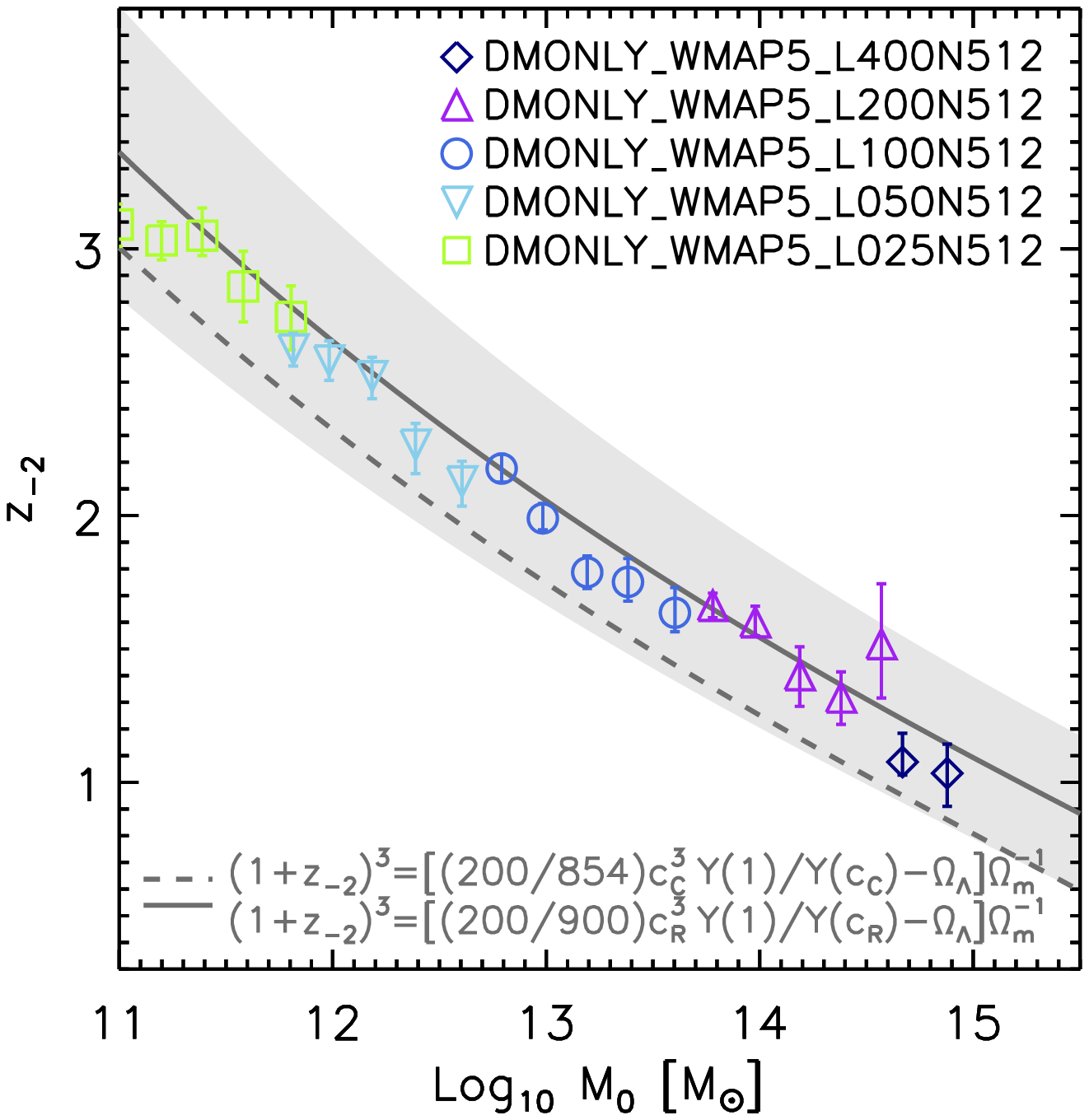}}
 \subfloat{\includegraphics[angle=0,width=0.48\textwidth]{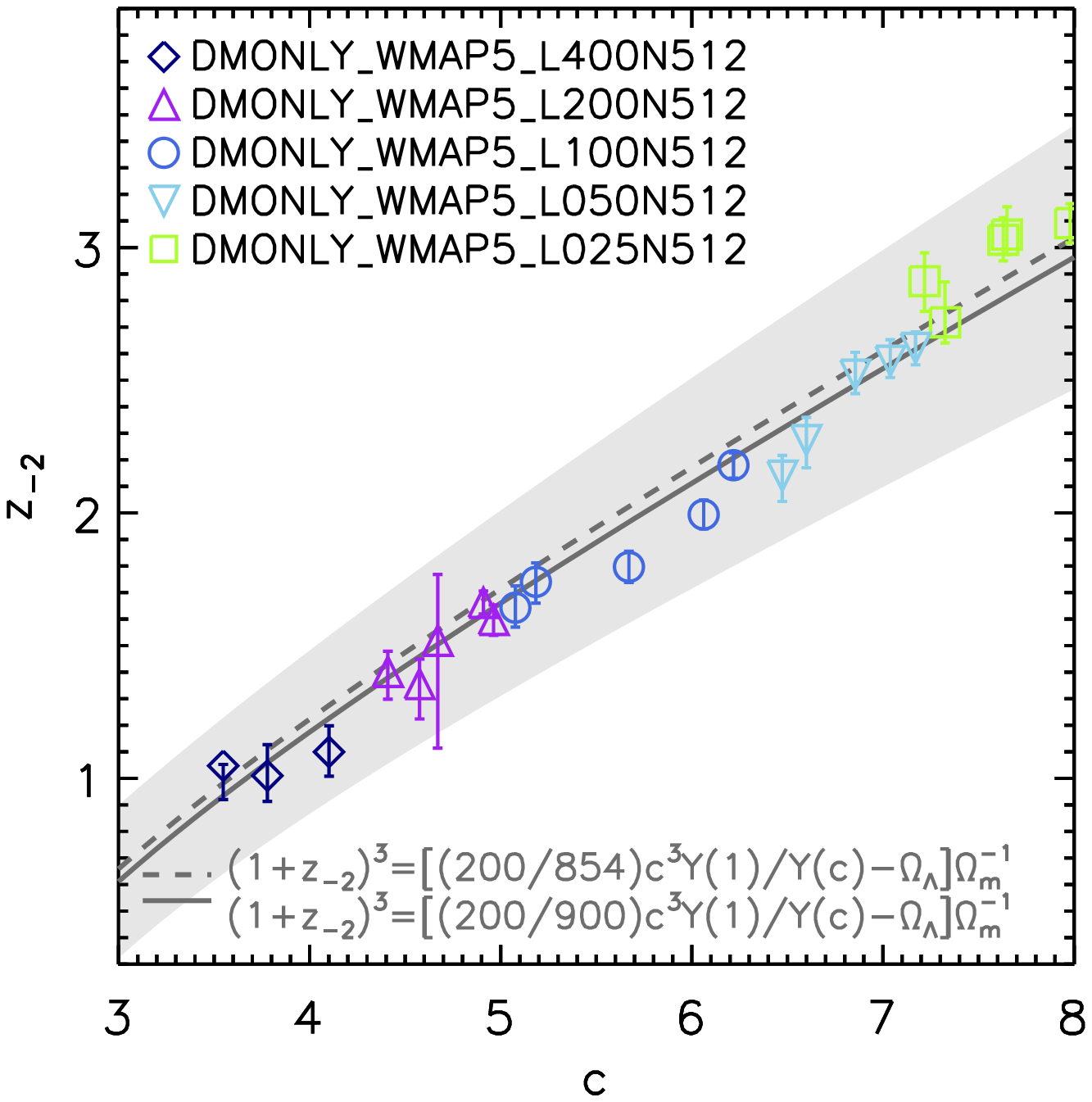}}
\caption{Relation between formation redshift, $z_{-2}$, and halo concentration, $c$ (left panel), and between formation redshift and $z=0$ halo mass, $M_{200}$ (right panel). The different symbols correspond to the median values of the relaxed sample and the error bars to $1\sigma$ confidence limits. The solid line in the left panel is not a fit but a prediction of the $z_{-2}-c$ relation for relaxed halos given by eq.~(\ref{zf-c}). Similarly, the dashed line is the prediction for the complete sample of halos, assuming a constant of proportionality between $\langle\rho\rangle(<r_{-2})$ and $\rho_{\rm{crit}}$ of $854$, rather than the value of $900$ used for the relaxed sample. The grey area shows the scatter in $z_{-2}$ plotted in Fig. \ref{scatter_plot} (right panel). Similarly, the solid line in the right panel is a prediction of the $z_{-2}-M_{200}$ relation given by eqs.~(\ref{zf-c}) and (\ref{c-m}). The dashed line also shows the $z_{-2}-M_{200}$ relation assuming $\langle\rho\rangle(<r_{-2})=854\rho_{\rm{crit}}$ and the concentration-mass relation calculated using the complete sample.}
\label{zformation}
\end{figure*}

\subsection{Relation between halo formation time and concentration from numerical simulations}

We now study the relation between $\rho_{\rm{crit}}(z_{-2})$ and $\langle\rho\rangle(<r_{-2})$ using a set of DMONLY cosmological simulations from the OWLS project that adopt the WMAP-5 cosmology. We begin by considering two samples of halos. Our complete sample contains all halos that satisfy our resolution criteria while our `relaxed' sample retains only those halos for which the separation between the most bound particle and the centre of mass of the Friends-of-Friends (FoF) halo is smaller than
$0.07R_{\rm{vir}}$, where $R_{\rm{vir}}$ is the radius for which the mean internal density is $\Delta$ (as given by \citealt{BryanNorman}) times the critical density. \citet{Neto} found that this simple criterion resulted in the removal of the vast majority of unrelaxed halos and as such we do not use their additional criteria. At $z=0$, our complete sample contains 2831 halos, while our relaxed sample is reduced to 2387 ($84\%$). 

To compute the mean inner density within the scale radius, $\langle\rho\rangle(<r_{-2})$, we need to fit the NFW density profile to each individual halo. We begin by fitting NFW profiles to all halos at $z=0$ that contain at least $10^{4}$ dark matter particles within the virial radius. For each halo, all particles in the range $-1.25\le \log_{10}(r/r_{200})\le 0$, where $r_{200}$ is the virial radius, are binned radially in equally spaced logarithmic bins of size $\Delta\log_{10}r=0.078$. The density profile is then fit to these bins by performing a least square minimization of the difference between the logarithmic densities of the model and the data, using equal weighting. The corresponding mean enclosed mass, $M_{r}(r_{-2})$, and mean inner density at $r_{-2}$, $\langle\rho\rangle(<r_{-2})$, are found by interpolating along the cumulative mass and density profiles (measured while fitting the NFW profile) from $r=0$ to $r_{-2}=r_{200}/c$, where $c$ is the concentration from the NFW fit. Then we follow the mass history of these halos through the snapshots, and interpolate to determine the redshift $z_{-2}$ at which $M_{z}=M_{r}(r_{-2})$. 

We perform a least-square minimization of the quantity $\Delta^{2} = \frac{1}{N}\sum_{i=1}^{N}[\langle\rho_{i}\rangle(\rho_{\rm{crit},i})-f(\rho_{\rm{crit},i},A)]$, to obtain the constant of proportionality, $A$. We find $\langle\rho\rangle(<r_{-2}) = (900\pm50)\rho_{\rm{crit}}(z_{-2})$ for the relaxed sample, and $\langle\rho\rangle(<r_{-2}) = (854\pm47)\rho_{\rm{crit}}(z_{-2})$ for the complete sample. The $1\sigma$ error was obtained from the least squares fit. For comparison, \cite{Ludlow14} found a constant value of $853$ for their relaxed sample of halos and the WMAP-1 cosmology. Fig. \ref{density_relation} shows the relation between the mean inner density at $z=0$ and the critical density of the universe at redshift $z_{-2}$ for various DMONLY simulations. Each dot in this panel corresponds to an individual halo from the complete sample in the DMONLY$_{-}$WMAP5 simulations that have box sizes of 400$\hMpc$, 200$\hMpc$, 100$\hMpc$, 50$\hMpc$ and 25$\hMpc$. The $\langle\rho\rangle(<r_{-2})$-$\rho_{\rm{crit}}(z_{-2})$ values are coloured by mass according to the colour bar at the top of the plot. The black (red) star symbols show the mean values of the complete (relaxed) sample in logarithmic mass bins of width $\delta\log_{10}M=0.2$. As expected when unrelaxed halos are discarded (e.g. Duffy et al. 2008), the relaxed sample contains on average slightly higher concentrations (by a factor of 1.16) and so higher formation times (by a factor of 1.1).

In Fig. \ref{density_relation} the best-fit to the data points from the relaxed sample is shown by the solid line, while the dashed line is the fit to the complete sample. The $\rho_{\rm{crit}}(z_{-2})-\langle\rho\rangle(<r_{-2})$ correlation clearly shows that halos that collapsed earlier have denser cores at $z=0$. Using the mean inner density-critical density relation for the relaxed sample, 

\begin{eqnarray}\label{rho-crit}
\langle\rho\rangle(<r_{-2}) &=& (900\pm 50)\rho_{\rm{crit}}(z_{-2}).\\\nonumber
\end{eqnarray}

\noindent We replace $\langle\rho\rangle(<r_{-2})$ by eq.~(\ref{rho-2}) and calculate the dependence of formation redshift on concentration,

\begin{eqnarray}\label{zf-c}
(1+z_{-2})^{3} &=& \frac{200}{900}\frac{c^{3}}{\Omega_{m}}\frac{Y(1)}{Y(c)}-\frac{\Omega_{\Lambda}}{\Omega_{m}}.
\end{eqnarray}

\noindent This last expression is tested in Fig. \ref{zformation} (left panel) where we plot the median formation redshift as a function of concentration using different symbols for different sets of simulations from the OWLS project. The symbols correspond to the median values of the relaxed sample, and the error bars indicate $1\sigma$ confidence limits. The grey solid line shows the $z_{-2}-c$ relation given by eq.~(\ref{zf-c}), whereas the grey dashed line shows the same relation assuming a constant of $854$ instead of $900$ (as obtained for the complete sample). Similarly, using the Duffy et al. (2008) concentration-mass relation we obtain the formation redshift as a function of halo mass at $z=0$ (right panel of Fig. \ref{zformation}). It is important to note that the $z_{-2}-c$ and $z_{-2}-M$ relations are valid in the halo mass range $10^{11}-10^{15}\hMsol$, at lower masses the concentration-mass relation begins to deviate from power laws (\citealt{Ludlow14}).

In Appendix \ref{Scatter_analysis} we analyse the scatter in the formation time$-$mass relation and show that it correlates with the scatter in the concentration$-$mass relation. Thus concluding that the scatter in formation time determines the scatter in the concentration. Also, we investigate how the scatter in halo mass history drives the scatter in formation time.


\subsection{The mass history}\label{MAH_semianalytic}

\begin{figure} 
\begin{center}
\includegraphics[angle=0,width=0.48\textwidth]{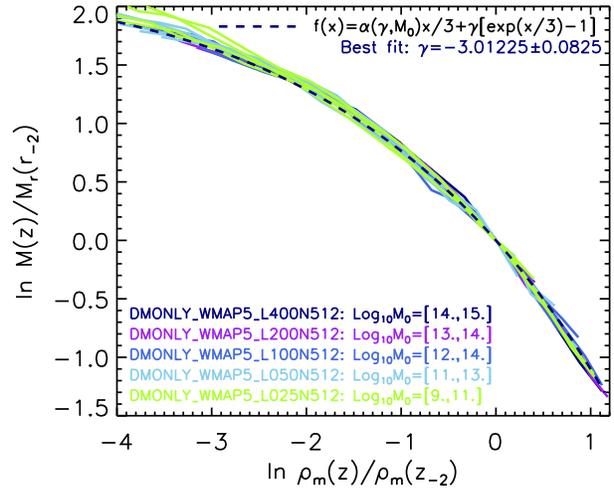}%
\caption{Mass histories of halos, obtained from different DMONLY$_{-}$WMAP5 simulations, as indicated by the colours. The bottom left legends indicate the halo mass range at $z=0$, selected from each simulation. For example, we selected halos between $10^{9}-10^{11}\rm{M}_{\sun}$ from the DMONLY$_{-}$WMAP5$_{-}$L025N512 simulation, divided them in equally spaced logarithmic bins of size $\Delta\log_{10} M=0.2$, and calculated the median mass histories. The different curves show the median mass history of the main progenitors, normalized to the median enclosed mass of the main progenitors at $z=0$, $M_{r}(r_{-2})$. The mass histories are plotted as a function of the mean background density of the universe, scaled to the mean background density at $z_{-2}$. The blue dashed line is a fit of expression (\ref{function}) to the different mass history curves. The median value of the only adjustable parameter, $\gamma$, is indicated in the top-right part of the plot.}
\label{unico_fit}
\end{center}
\end{figure}

Fig. \ref{unico_fit} shows the mass accretion history of halos in different mass bins as a function of the mean background density. The mass histories are scaled to $M_{r}(r_{-2})$ and the mean background densities are scaled to $\rho_{\rm{m}}(z_{-2})=\rho_{\rm{crit,0}}\Omega_{\rm{m}}(1+z_{-2})^{3}$. The figure shows that all halo mass histories look alike. This is in agreement with \citet{Ludlow}, who found that the mass accretion history, expressed in terms of the critical density of the Universe, $M(\rho_{\rm{crit}}(z))$, resembles that of the enclosed NFW density profile, $M(\langle\rho\rangle(<r))$. The similarity in the shapes between $M(\rho_{\rm{crit}}(z))$ and $M(\langle\rho\rangle(<r))$ is still not clear, but it suggests that the physically motivated form $M(z)=M_{0}(1+z)^{\alpha}e^{\beta z}$, which is a result of rapid growth in the matter dominated epoch followed by a slow growth in the dark energy epoch, produces the double power-law of the NFW profile (see e.g. \citealt{Lu2006}). We use this feature to find a functional form that describes this unique universal curve in order to obtain an empirical expression for the mass accretion history at all redshifts and halo masses. 

We are motivated by the extended Press-Schechter analysis of halo mass histories presented in Paper I. In that work, we showed through analytic calculations, that when halo mass histories are described by a power-law times an exponential, 

\begin{equation}\label{McBride}
M(z)=M_{0}(1+z)^{\alpha}e^{\beta z},
\end{equation}

\noindent and that the parameters $\alpha$ and $\beta$ are connected via the power spectrum of density fluctuations. In this work however, we aim to relate halo structure to the mass accretion history. We therefore first determine the correlation between the parameters $\alpha$ and $\beta$ and concentration. To this end, we first find the $\alpha-\beta$ relation that results from the formation redshift definition discussed in the previous section. Thus, we evaluate the halo mass at $z_{-2}$,

\begin{eqnarray}\label{McBride2}
M_{z}(z_{-2}) = M_{r}(r_{-2}) &=& M_{z}(z=0)(1+z_{-2})^{\alpha}e^{\beta z_{-2}}.
\end{eqnarray}
\noindent Taking the natural logarithm, we obtain,
\begin{eqnarray}
\ln\left(\frac{M_{r}(r_{-2})}{M_{z}(z=0)}\right) &=& \alpha\ln(1+z_{-2})+\beta z_{-2},\\\nonumber
\end{eqnarray}
\noindent and hence
\begin{eqnarray}
\alpha &=& \frac{\ln\left(\frac{M_{r}(r_{-2})}{M_{z}(z=0)}\right)-\beta z_{-2}}{\ln(1+z_{-2})}.
\end{eqnarray}

\noindent From this last equation we see that $\alpha$ can be written as a function of $\beta$, $M_{r}(r_{-2})$, $M_{z}(z=0)$ and $z_{-2}$. However, as $M_{r}(r_{-2})$ is a function of concentration and virial mass (see eq.~\ref{M2}), we can write $\alpha$ in terms of $\beta$, concentration and $z_{-2}$,

\begin{eqnarray}\label{alpha2}
\alpha &=& \frac{\ln(Y(1)/Y(c))-\beta z_{-2}}{\ln(1+z_{-2})}.
\end{eqnarray}

The next step is to find an expression for $\beta$. We find $\beta(z_{-2})$ by fitting eq.~(\ref{McBride}) to all the data points plotted in Fig.~\ref{unico_fit}. We now need to express $M(z)$ (eq.~\ref{McBride}) as a function of the mean background density. To do this, we replace $(1+z)$ by $(\rho_{\rm{m}}(z)/\rho_{\rm{crit},0}/\Omega_{\rm{m}})^{1/3}$ and divide both sides of eq.~(\ref{McBride}) by $M_{r}(r_{-2})$, yielding

\begin{eqnarray}\nonumber
\frac{M_{z}(z)}{M_{r}(r_{-2})} &=&  \frac{M_{z}(z=0)}{M_{r}(r_{-2})}\left(\frac{\rho_{\rm{m}}(z)}{\Omega_{\rm{m}}\rho_{\rm{crit,0}}}\right)^{\alpha/3}\\
& \times &{\rm{exp}}\left(\beta\left[\left(\frac{\rho_{{\rm{m}}}(z)}{\Omega_{{\rm{m}}}\rho_{{\rm{crit,0}}}}\right)^{1/3}-1\right]\right).
\end{eqnarray}

\noindent Multiplying both denominators and numerators by $\rho_{\rm{m}}(z_{-2})$, we get, after rearranging,

\begin{eqnarray}\nonumber
\frac{M_{z}(z)}{M_{r}(r_{-2})} &=&  \frac{M_{z}(z=0)}{M_{r}(r_{-2})}\left(\frac{\rho_{\rm{m}}(z_{-2})}{\Omega_{\rm{m}}\rho_{\rm{crit,0}}}\right)^{\alpha/3}\left(\frac{\rho_{\rm{m}}(z)}{\rho_{\rm{m}}(z_{-2})}\right)^{\alpha/3}\\\nonumber
& \times &  {\rm{exp}}\left(\beta\left[\left(\frac{\rho_{\rm{m}}(z_{-2})}{\Omega_{\rm{m}}\rho_{\rm{crit,0}}}\right)^{1/3}-1\right]\right)\\\label{expression}
&\times & {\rm{exp}}\left(\gamma\left[\left(\frac{\rho_{\rm{m}}(z)}{\rho_{\rm{m}}(z_{-2})}\right)^{1/3}-1\right]\right),\\\nonumber
\end{eqnarray} 

\noindent where, we have defined $\gamma\equiv\beta(\rho_{\rm{m}}(z_{-2})/\Omega_{\rm{m}}/\rho_{\rm{crit,0}})^{1/3}=\beta(1+z_{-2})$. The term $\frac{M_{z}(z=0)}{M_{r}(r_{-2})}\left(\frac{\rho_{\rm{m}}(z_{-2})}{\Omega_{\rm{m}}\rho_{\rm{crit,0}}}\right)^{\alpha/3}{\rm{exp}}\left(\beta\left[\left(\frac{\rho_{\rm{m}}(z_{-2})}{\Omega_{\rm{m}}\rho_{\rm{crit,0}}}\right)^{1/3}-1\right]\right)$ in eq.~(\ref{expression}) is equal to unity, which can be seen by replacing $\rho_{\rm{m}}(z_{-2})/\Omega_{\rm{m}}\rho_{\rm{crit,0}}=(1+z_{-2})^{3}$ and comparing with eq. (\ref{McBride2}). Hence eq. (\ref{expression}) becomes

\begin{eqnarray}\nonumber
\frac{M_{z}(z)}{M_{r}(r_{-2})} &=&  \left(\frac{\rho_{\rm{m}}(z)}{\rho_{\rm{m}}(z_{-2})}\right)^{\alpha/3}\\\label{expression2}
&\times & {\rm{exp}}\left(\gamma\left[\left(\frac{\rho_{\rm{m}}(z)}{\rho_{\rm{m}}(z_{-2})}\right)^{1/3}-1\right]\right).\\\nonumber
\end{eqnarray}

Thus, based on eq.~(\ref{expression2}), the functional form to fit the mass accretion histories from the simulations can be written as

\begin{eqnarray}\label{function}
f(\tilde{z},\gamma) &=& \alpha(z_{-2},c,\gamma)\tilde{z}/3+\gamma(e^{\tilde{z}/3}-1),\\\nonumber
\end{eqnarray}

\noindent where $f(\tilde{z},\gamma)=\ln\left(\frac{M_{z}(z)}{M_{r}(r_{-2})}\right)$ and $\tilde{z}=\ln\left[\frac{\rho_{\rm{m}}(z)}{\rho_{\rm{m}}(z_{-2})}\right]$. From eq. (\ref{alpha2}) we see that the parameter $\alpha$ is a function of $z_{-2}$, $c$ and $\gamma$,

\begin{eqnarray}\label{alpha3}
\alpha &=& \frac{\ln(Y(1)/Y(c))-\gamma z_{-2}/(1+z_{-2})}{\ln(1+z_{-2})}.
\end{eqnarray}

\noindent Therefore, $\gamma$ is now the only adjustable parameter. We perform a $\chi^{2}$-like minimization of the quantity

\begin{equation}
\Delta^{2}=\frac{1}{N}\sum^{N}_{i=1}[\log_{10}(M_{z}(z_{i})/M_{r}(r_{-2}))-f(\tilde{z}_{i},\gamma)]^{2},
\end{equation}

\noindent and find the value of $\gamma$ that best fits all halo accretion histories. The sum in the $\chi^{2}$-like minimization is over the $N$ available simulation output redshifts at $z_{i}(i=1,N)$, with $\tilde{z_{i}}=\ln\left[\frac{\rho_{\rm{crit,0}}\Omega_{\rm{m}}(1+z_{i})^{3}}{\rho_{\rm{m}}(z_{-2})}\right]$. 

\begin{figure*}
\centering
\includegraphics[angle=0,width=0.9\textwidth]{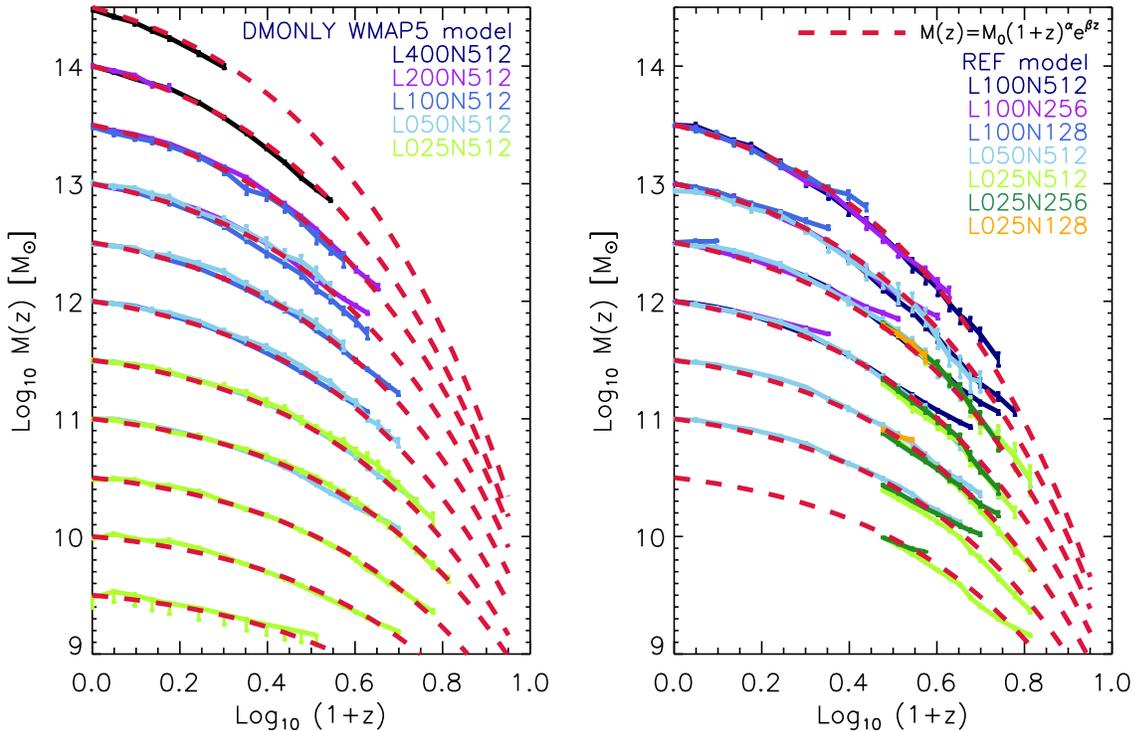}%
\caption{Mass histories of all halos from simulations DMONLY$_{-}$WMAP5 (left panel) and REF (right panel). Halo masses are binned in equally spaced logarithmic bins of size $\Delta \log_{10}M=0.5$. The mass histories are computed by calculating the median value of the halo masses from the merger tree at a given output redshift, the error bars correspond to 1$\sigma$. The different colour lines show the mass histories of halos from different simulations as indicated in the legends, while the red dashed curves correspond to the mass histories predicted by eqs.~(\ref{McBride}), (\ref{alpha3}) and (\ref{beta}).}
\label{comparison}
\end{figure*}

Fig. \ref{unico_fit} shows halo mass histories (with $M_{z}(z)$ scaled to $M_{r}(r_{-2})$) for our complete halo sample as a function of the mean background density [$\rho_{\rm{m}}(z)$ scaled to $\rho_{\rm{m}}(z_{-2})$]. The blue dashed line is the fit of expression (\ref{function}) to all the mass history curves included in the figure. Here, the only adjustable parameter is $\gamma$. We obtained $\gamma=-3.01\pm 0.08$, yielding
\begin{eqnarray}\label{beta}
\beta = -3(\rho_{\rm{m}}(z_{-2})/\Omega_{\rm{m}}/\rho_{\rm{crit,0}})^{-1/3} = -3/(1+z_{-2}).
\end{eqnarray}

Fig. \ref{unico_fit} shows that the halo mass histories have a characteristic shape consisting of a rapid growth at early times, followed by a slower growth at late times. The change from rapid to slow accretion corresponds to the transition between the mass and dark energy dominated eras (see Paper I), and depends on the parameter $\beta$ in the exponential (as can be seen from eq.~\ref{McBride}). The dependence of $\beta$ on the formation redshift is given in eq.~(\ref{beta}), which shows that a more recent formation time, and hence a larger halo mass, results in a larger value of $\beta$, and so a steeper halo mass history curve. This last point can be seen in Fig. \ref{comparison} from the mass histories of halos in different mass bins (coloured lines shown in the panels). The panel on the left shows the mass history curves from the DMONLY simulation outputs (coloured lines in the background), and the mass histories predicted by eqs.~(\ref{McBride}), (\ref{alpha2}) and (\ref{beta}) (red dashed lines). From these panels we see that, (i) the mass history formula works remarkably well when compared with the simulation, and (ii) the larger the mass of a halo at $z=0$, the steeper the mass history curve at early times. In contrast, the mass history of low-mass halos is essentially governed by the power law at late times. 

The halo mass histories plotted in Fig. \ref{comparison} come from the complete sample of halos (relaxed and unrelaxed). We found no significant difference in mass growth when only relaxed halos are considered. We therefore conclude that the fact that a halo is unrelaxed at a particular redshift does not affect its halo mass history, provided the concentration$-$mass relation fit from the relaxed halo sample is used. This is an interesting result because while deriving the semi-analytic model of halo mass history, we assumed that the halo density profile is described by the NFW profile at all times. Therefore while the NFW is not a good fit for the density profile of unrelaxed halos (\citealt{Neto}), our semi-analytic model (based on NFW profiles) is a good fit all halos (relaxed and unrelaxed).

The right panel of Fig. \ref{comparison} shows halo mass histories from the REF hydrodynamical simulations. We compute the halo mass as the total mass (gas and dark matter) within the virial radius ($r_{200}$). We find that the inclusion of baryons steepens the mass histories at high redshift, therefore the best description of $M(z)$ is given by eqs.~(\ref{McBride}), (\ref{alpha2}), (\ref{beta}), and the concentration-mass relation from the complete sample of halos, ${c=5.74(M/2\times 10^{12}h^{-1}\rm{M}_{\sun})^{-0.097}}$. 

\subsection{The mass accretion rate}

The accretion of gas and dark matter from the intergalactic medium is a fundamental driver of both, the evolution of dark matter halos and the formation of galaxies within them. For that reason, developing a theoretical model for the mass accretion rate is the basis for analytic and semi-analytic models that study galaxy formation and evolution. In this section we look for a suitable expression for the mean accretion rate of dark matter halos. To achieve this, we take the derivative of the semi-analytic mass history model, $M(z)$, given by eq.~(\ref{McBride}) with respect to time and replace $dz/dt$ by $-H_{0}[\Omega_{\rm{m}}(1+z)^{5}+\Omega_{\Lambda}(1+z)^{2}]^{1/2}$, yielding

\begin{eqnarray}\nonumber
\frac{dM}{dt} &=& 71.6 {\rm{M}}_{\sun}{\rm{yr}}^{-1} M_{12}h_{0.7}[-\alpha-\beta(1+z)]\\\label{accretion_rate}
&& \times[\Omega_{\rm{m}}(1+z)^{3}+\Omega_{\Lambda}]^{1/2},
\end{eqnarray}

\noindent where $h_{0.7}=h/0.7$, $M_{12}=M/10^{12}\rm{M}_{\sun}$ and $\alpha$ and $\beta$ are given by eqs.~(\ref{alpha2}) and (\ref{beta}), respectively. As shown in the previous section, the parameters $\alpha$ and $\beta$ depend on halo mass (through the formation time dependence). We find that this mass dependence is crucial for obtaining an accurate description for the mass history (as shown in Fig. \ref{comparison}). However, the factor of $2$ ($3$) change for $\alpha$ ($\beta$) between halo masses of $10^{8}$ and $10^{14}\rm{M}_{\sun}$ is not significant when calculating the accretion rate. Therefore, we provide an approximation for the {\it{mean}} mass accretion rate as a function of redshift and halo mass, by averaging $\alpha$ and $\beta$ over halo mass, yielding $\langle\alpha\rangle$=0.24, $\langle\beta\rangle$=-0.75, and

\begin{eqnarray}\nonumber
\langle\frac{dM}{dt}\rangle &=& 71.6 {\rm{M}}_{\sun}{\rm{yr}}^{-1} M_{12}h_{0.7}\\\label{mean_accr}
&& \times[-0.24+0.75(1+z)][\Omega_{\rm{m}}(1+z)^{3}+\Omega_{\Lambda}]^{1/2}.
\end{eqnarray}

\begin{figure} 
\begin{center}
\includegraphics[angle=0,width=0.48\textwidth]{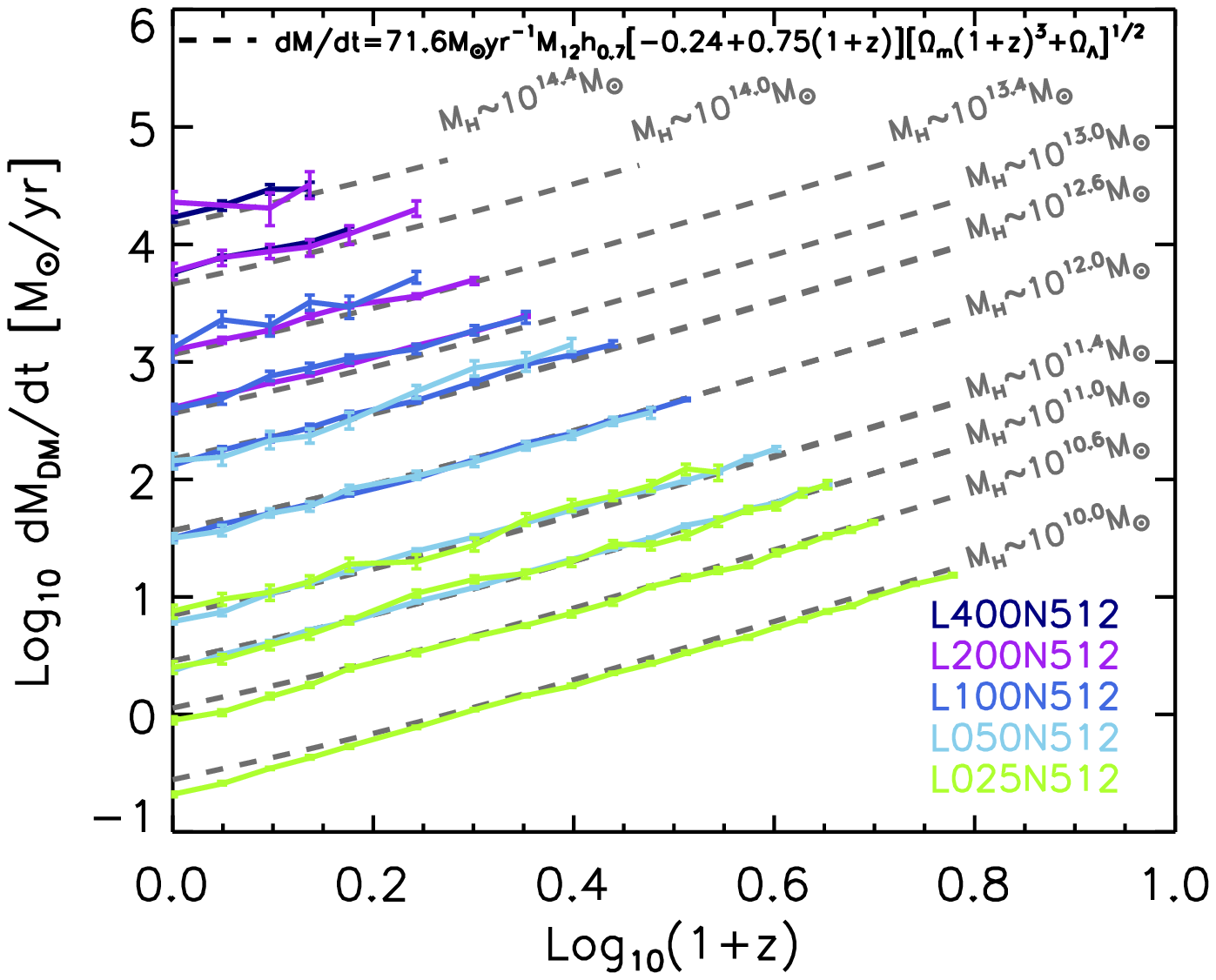}\\
\includegraphics[angle=0,width=0.48\textwidth]{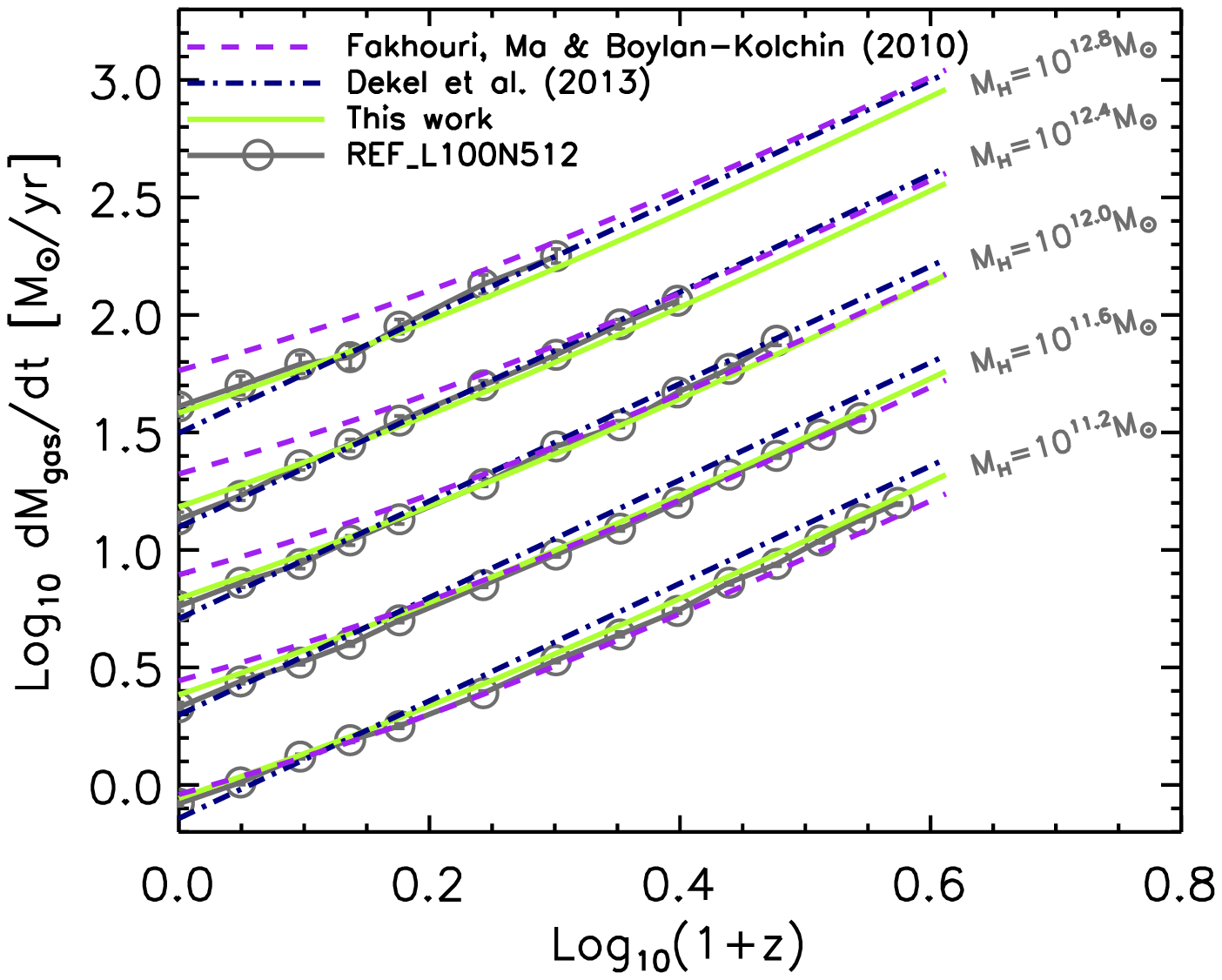}%
\caption{Mean accretion rate of dark matter (top panel) and gas (bottom panel) as a function of redshift for different halo masses. {\it{Top panel}}: the accretion rate obtained from the simulation outputs up to the redshift where the halo mass histories are converged. Grey dashed lines show the accretion rate estimated using eq.~(\ref{mean_accr}). {\it{Bottom panel}}: gas accretion rate obtained from the REF$_{-}$L100N512 simulation (grey circles), from $\Omega_{\rm{b}}/\Omega_{\rm{m}}$ times eq.~(\ref{mean_accr}) (green solid line), and from various fitting formulae taken from the literature.}   
\label{MAH}
\end{center}
\end{figure}

Fig. \ref{MAH} (top panel) compares the median dark matter accretion rate for different halo masses as a function of redshift (solid lines) to the mean accretion rate given by eq.~(\ref{mean_accr}) (grey dashed lines). From the merger trees of the main halos, we compute the mass growth rate of a halo of a given mass. We do this by following the main branch of the tree and computing $dM/dt = (M(z_{1})-M(z_{2}))/\Delta t$, where $z_{1}<z_{2}$, $M(z_{1})$ is the descendant halo mass at time $t$ and $M(z_{2})$ is the most massive progenitor at time $t-\Delta t$. The median value of $dM/dt$ for the complete set of resolved halos is then plotted for different constant halo masses. We find very good agreement between the simulation outputs and the analytic estimate given by eq.~(\ref{mean_accr}). As expected, the larger the halo mass, the larger the dark matter accretion rate.

\subsubsection{Baryonic accretion}

Next, we estimate the gas accretion rate and compare our model with similar fitting formulae proposed by \citet{Fakhouri} and \citet{Dekel13}. The bottom panel of Fig. \ref{MAH} shows the gas accretion rate as a function redshift for a range of halo masses ($\log_{10}M/\rm{M}_{\sun}=11.2-12.8$). The grey circles correspond to the gas accretion rate measured	 in REF$_{-}$L100N512. In this case we compute the total mass growth ($M=M_{\rm{gas}}+M_{\rm{DM}}$) from the merger trees, and then estimate the gas accretion rate by multiplying the total accretion rate by the universal baryon fraction $f_{b}=\Omega_{\rm{b}}/\Omega_{\rm{m}}$. The green solid line corresponds to our gas accretion rate model (given by $\Omega_{\rm{b}}/\Omega_{\rm{m}}$ times eq.~\ref{mean_accr}). The blue dot-dashed line is the gas accretion rate proposed by \citet{Dekel13} ($dM_{\rm{b}}/dt = 30{\rm{M}}_{\sun}{\rm{yr}}^{-1}f_{b}M_{12}(1+z)^{5/2}$), who derived the baryonic inflow on to a halo $dM_{\rm{b}}/dt$ from the averaged growth rate of halo mass through mergers and smooth accretion based on the EPS theory of gravitational clustering (\citealt{Neistein,Neistein08}). Lastly, we compare our model with the accretion rate formula from \citet{Fakhouri} [$dM_{\rm{b}}/dt = 46.1{\rm{M}}_{\sun}{\rm{yr}}^{-1}f_{\rm{b}}M_{12}^{1.1}(1+1.11z)(\Omega_{\rm{m}}(1+z)^{3}+\Omega_{\Lambda})^{1/2}$], plotted as the purple dashed line. \citet{Fakhouri} constructed merger trees of dark matter halos and quantified their merger rates and mass growth rate using the Millennium and Millennium II simulations. They defined the halo mass as the sum of the masses of all subhalos within a FoF halo. We see that our accretion rate model is in excellent agreement with the formulae from \citet{Fakhouri} and \citet{Dekel13}. We find that the \citet{Fakhouri} formula generally overpredicts the gas accretion rate in the low-redshift regime (e.g. it overpredicts it by a factor of 1.4 at $z=0$ for a $10^{12}\rm{M}_{\sun}$ mass halo). The \citet{Dekel13} formula underpredicts (overpredicts) the gas accretion rate in the low- (high-) redshift regime for halos with masses larger (lower) than $10^{12}\rm{M}_{\sun}$.

\subsection{Dependence on cosmology and mass definition}\label{Discussion}

We have developed a semi-analytic model that relates the inner structure of a halo at redshift zero to its mass history. The model adopts the NFW profile, computes the mean inner density within the scale radius, and relates this to the critical density of the universe at the redshift where the halo virial mass equals the mass enclosed within $r_{-2}$. This relation enables us to find the formation redshift-halo mass dependence and to derive a one-parameter model for the halo mass history. In this section we consider the effects of cosmology and mass definition on the semi-analytic model.

\subsubsection{Cosmology dependence}\label{cosmo_sec}

The adopted cosmological parameters affect the mean inner halo densities, concentrations, formation redshifts and halo mass histories. To investigate the dependence of halo mass histories on cosmology, we have run a set of dark matter only simulations with different cosmologies. Table \ref{cosmo} lists the sets of cosmological parameters adopted by the different simulations. Specifically, we assume values for the cosmological parameters derived from measurements of the cosmic microwave background by the WMAP and the {\it{Planck}} missions (\citealt{Spergel03,Spergel07, Komatsu,Hinshaw,Planck}). 

It has been shown that halos that formed earlier are more concentrated (\citealt{NFW97,Bullock,Eke, Kuhlen,Maccio07,Neto}). \citet{Maccio08} explored the dependence of halo concentration on the adopted cosmological model for field galaxies. They found that dwarf-scale field halos are more concentrated by a factor of 1.55 in WMAP1 compared to WMAP3, and by a factor of 1.29 for cluster-sized halos. This reflects the fact that halos of a fixed $z=0$ mass assemble earlier in a universe with higher $\Omega_{\rm{m}}$, higher $\sigma_{8}$ and/or higher $n_{s}$.

The halo formation redshift can be related to the power at the corresponding mass scale, and therefore depends on both $\sigma_{8}$ and $n_{s}$. The parameter $\sigma_{8}$ sets the power at a scale of $8\hMpc$, which corresponds to a mass of about $1.53\times 10^{14}h^{-1}\rm{M}_{\sun}(\Omega_{\rm{m}}/\Omega_{\rm{m,WMAP5}})$, and a wavenumber of $k_{8}$. This last quantity is given by the relation ${M=(4\pi\rho_{\rm{m}}/3)(2\pi/k)^{3}}$. For a power-law spectrum ${P(k)\propto k^{n}}$, the variance can be written as ${\sigma^{2}(k)/\sigma^{2}_{8}=(k/k_{8})^{n+3}}$. Therefore, the change in $\sigma$ between WMAP5 and WMAP1 for a given halo mass that corresponds to a wavenumber $k$ is

\begin{equation}
\frac{\sigma_{\rm{WMAP1}}(k)}{\sigma_{\rm{WMAP5}}(k)}=\frac{\sigma_{8,\rm{WMAP1}}}{\sigma_{8,\rm{WMAP5}}}\left(\frac{k}{k_{8}}\right)^{(n_{\rm{s,WMAP1}}-n_{\rm{s,WMAP5}})/2}.
\end{equation}

\noindent A halo mass of $10^{12}\rm{M}_{\sun}$ corresponds to a wavenumber of $k_{1.3}\sim 6k_{8}$. The total change in the mean power spectrum at this mass scale is ${\frac{\sigma_{\rm{WMAP1}}(k_{1.3})}{\sigma_{\rm{WMAP5}}(k_{1.3})}=1.27}$. This is proportional to the change of the formation redshift,

\begin{equation}
(1+z_{\rm{f,WMAP1}}) = 1.27(1+z_{\rm{f,WMAP5}}).
\end{equation}

Next, we test how this change affects the halo mass history. We showed that the mass history profile is well described by the expression ${M(z)=M_{z}(z=0)(1+z)^{\alpha}e^{\beta z}}$, where $\alpha$ and $\beta$ both depend on the formation redshift. In the mass history model presented in Section 4 there are two best-fitting parameters that can be cosmology dependent. One is the constant value $A=900$ in eq.~(\ref{rho-crit}) that relates the mean inner density to the critical density at $z_{-2}$, and the other is the constant value $\gamma=-3$ in eq.~(\ref{beta}) that defines the $\beta$ parameter.

To investigate the cosmology dependence of $A$, we analyse the $\langle\rho\rangle(<r_{-2})-\rho_{\rm{crit}}(z_{-2})$ relation in the simulations with different cosmologies. We do the same as in Section 4.3. First we fit the NFW profile to dark matter halos to obtain $c$ and $r_{-2}$, and calculate the cumulative mass, $M_{-2}$, and density, $\langle\rho\rangle(<r_{-2})$, from $r = 0$ to $r = r_{-2}$. Then we follow the halo mass histories through the snapshots and interpolate to calculate $z_{-2}$, the redshift for which $M(z)$ is equal to $M_{-2}$. Finally, we obtain the best-fitting $\langle\rho\rangle(<r_{-2})-\rho_{\rm{crit}}(z_{-2})$ relation. We find that the parameter $A_{\rm{cosmo}}$, where cosmo is WMAP1, WMAP3, WMAP5, WMAP9 or Planck, changes with cosmology. We show this in the top panel of Fig.~\ref{cosmotest}. We find $A_{\rm{WMAP1}}=787\pm 52.25$, $A_{\rm{WMAP3}}=850\pm 39.60$, $A_{\rm{WMAP5}}=903\pm 48.63$, $A_{\rm{WMAP9}}=820\pm 51.03$ and  $A_{\rm{Planck}}=798\pm 43.73$. We do not find good agreement with \cite{Ludlow}, who found $A_{\rm{WMAP1}}=853$ for WMAP1 cosmology. This is due to the fact that we are only analysing the $\langle\rho\rangle(<r_{-2})-\rho_{\rm{crit}}(z_{-2})$ relation in the high-mass regime ($M=10^{12.8}-10^{13.8}\rm{M}_{\sun}$), due to the limitations of the box size ($L=100\hMpc$). With a more complete halo population, we may obtain better agreement. We conclude that the $A_{\rm{cosmo}}$ parameter depends on cosmology, at least for the halo mass ranges we are considering.

\begin{figure} 
\begin{center}
\includegraphics[angle=0,width=0.48\textwidth]{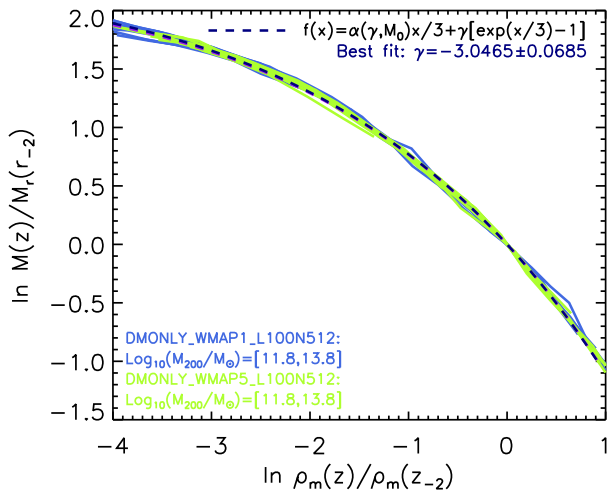}\\
\caption{Mass histories of halos, obtained from simulations with the WMAP1 cosmology (DMONLY$_{-}$WMAP1$_{-}$L100N512, blue solid lines) and the WMAP5 cosmology (DMONLY$_{-}$WMAP5$_{-}$L100N512, green solid lines). The curves show the median mass history of the main progenitors, normalized to the median enclosed mass, $M_{r}(r_{-2})$, of the main progenitors at $z=0$. The mass histories are plotted as a function of the mean background density of the universe, scaled to the mean background density at $z_{-2}$. The blue dashed line is a fit of expression (\ref{function}) to the different mass history curves. The median value of the only adjustable parameter, $\gamma$, is indicated in the top-right part of the plot. We find that $\gamma$ is insensitive to cosmology.}
\label{both_cosmo}
\end{center}
\end{figure}

\begin{figure} 
\begin{center}
\includegraphics[angle=0,width=0.46\textwidth]{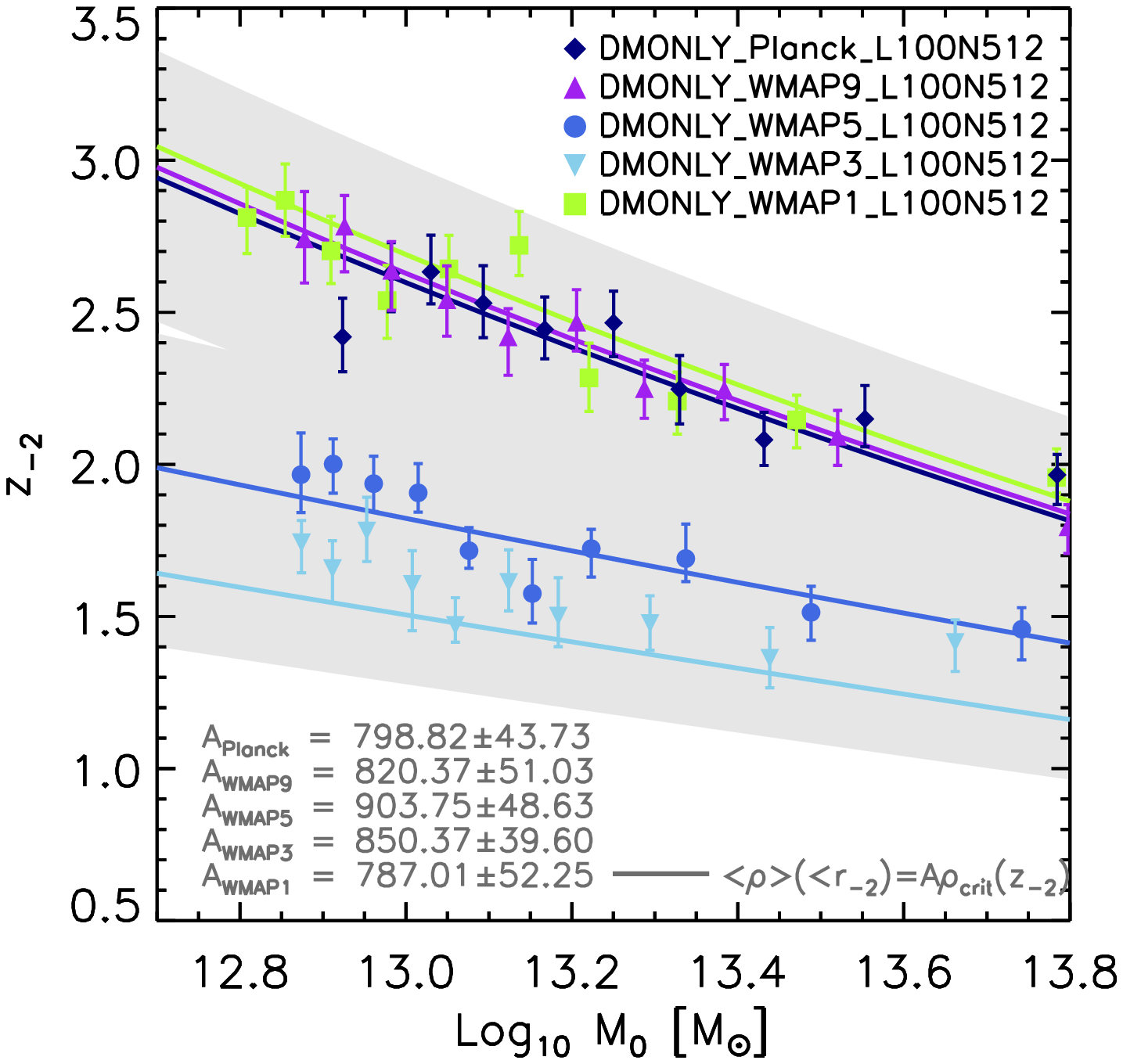}\\
\includegraphics[angle=0,width=0.48\textwidth]{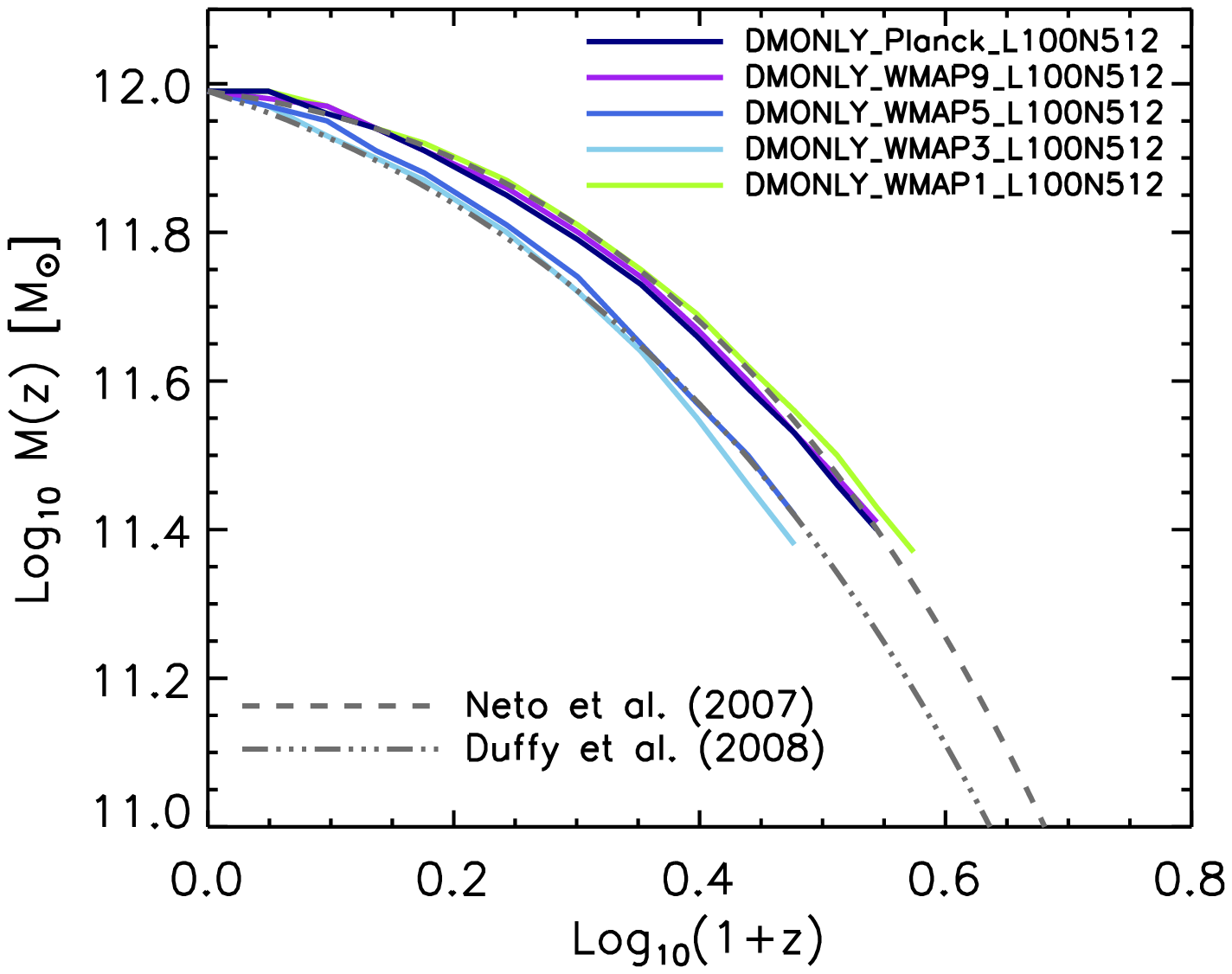}%
\caption{{\it{Top panel}}: relation between formation redshift ($z_{-2}$) and halo mass at $z=0$ ($M_{0}$). The different symbols correspond to median values, the error bars to $1\sigma$ confidence limits and the grey area to the scatter. These were computed from the dark matter only simulations that assumed a WMAP3 (light blue line), WMAP5 (blue line), WMAP1 (green line), WMAP9 (purple line) and Planck cosmology (dark blue line). The solid lines are not fits, but predictions of the $z_{-2}-M_{\rm{halo}}$ relation given by eq.~(\ref{zf-c}). We also indicated the different values of the constant of proportionality $A$ obtained by fitting the $\langle\rho_{-2}\rangle-\rho_{\rm{crit}}(z_{-2})$ relation. {\it{Bottom panel}}: halo mass history of a halo of $10^{12}\rm{M}_{\sun}$ at $z=0$ from DMONLY simulations with different cosmologies. The grey curves show that as long as a suitable concentration-mass relation is assumed for the cosmology under consideration, eqs.~(\ref{final_set1}), (\ref{final_set2}) and (\ref{final_set3}) give a good estimate of the mass history curve.}
\label{cosmotest}
\end{center}
\end{figure}

Next, we analyse how the change of the formation redshift due to cosmology affects $\beta$, defined as ${\beta=-3/(1+z_{\rm{f}})}$. We find from Fig.~\ref{both_cosmo} that the constant value, $-3$, is insensitive to cosmology. Fig.~\ref{both_cosmo} shows the same analysis as Fig.~\ref{unico_fit}, but for halo mass histories obtained from simulations with the WMAP1 and WMAP5 cosmology as indicated in the legends. We fit expression (\ref{function}) to the mass history curves from different cosmologies and obtained the same adjustable parameter $\gamma$. We therefore conclude that $\gamma=-3$ is insensitive to cosmology.

If we then consider that the change in $\beta$ between WMAP1 and WMAP5 is $\beta_{WMAP1}=-3/(1+z_{\rm{f,WMAP1}})=-3/1.27/(1+z_{\rm{f,WMAP5}})=\beta_{WMAP5}/1.27$, the change in the halo mass history between the WMAP5 and WMAP1 cosmologies, for a halo mass of $10^{12}\rm{M}_{\sun}$ at $z=0$, corresponds to

\begin{eqnarray}\nonumber
\log_{10}\frac{M(z)_{\rm{WMAP1}}}{M(z)_{\rm{WMAP5}}} &=& \log_{10}e^{(\beta_{\rm{WMAP1}}-\beta_{\rm{WMAP5}})z}\\\nonumber
& \approx & -0.12\beta_{\rm{WMAP1}}z,\\
& \approx & 0.1 z.
\end{eqnarray}

\noindent In the last step we replaced $\beta_{\rm{WMAP1}}$ by ${-3/(1+z_{\rm{f,WMAP1}})=-0.75}$ for a $10^{12}\rm{M}_{\sun}$ halo. We obtained $z_{\rm{f,WMAP1}}$ from the ${\langle\rho\rangle(<r_{-2})-\rho_{\rm{crit}}(z_{-2})}$ relation suitable for the WMAP1 cosmology (see top panel of Fig. \ref{cosmotest}) and the concentration-mass relation from \citet{Neto}. 

Next, we test the change in halo mass history. For example, if a halo had a mass of $10^{11.4}\rm{M}_{\sun}$ at $z=2$ in the WMAP5 cosmology, it would have had a mass of $10^{11.6}\rm{M}_{\sun}$ in the WMAP1 cosmology. The value of $\sigma_{8}$ has a particularly large effect at high redshift, because structure formation proceeds faster in the WMAP1 cosmology, as shown by the above expression. This last point can also be seen in the two panels of Fig. \ref{cosmotest}. The top panel is the same as the right panel of Fig. \ref{zformation}, and shows the formation redshift, $z_{-2}$, as a function of halo mass (obtained from simulations with different cosmologies). We see that there are large differences between the WMAP5 and WMAP1 cosmologies due to the changes in $\sigma_{8}$ and $n_{s}$. Interestingly, there is only a small difference between the Planck and WMAP1 cosmologies (in agreement with \citealt{Ludlow14} and \citealt{Dutton14}), and also between the Planck and WMAP9 cosmologies for which we found $(1+z_{\rm{f}})_{\rm{Planck}} = 1.01(1+z_{\rm{f}})_{\rm{WMAP9}}$ for a $10^{12}\rm{M}_{\sun}$ halo. The bottom panel of Fig. \ref{cosmotest} shows the mass history of $10^{12}\rm{M}_{\sun}$ halos at $z=0$ from DMONLY simulations with different cosmologies. As predicted, the change in mass history between the WMAP5 and WMAP1 cosmologies is $\log_{10}\frac{M(z)_{\rm{WMAP1}}}{M(z)_{\rm{WMAP5}}}\sim 0.1 z$, while little difference is found between the WMAP9 and Planck ($\Delta M(z)\sim 10^{-3}z$). 

We found that as long as a suitable concentration-mass relation and value for the $A$ parameter are assumed for the cosmology being considered, eqs.~(\ref{final_set1}), (\ref{final_set2}) and (\ref{final_set3}) provide a good estimate of the mass history curve. This can be seen in the bottom panel of Fig. \ref{cosmotest} by comparing the different halo mass histories. For the WMAP5 and WMAP3 cosmologies we assumed the concentration-mass relation found by \citet{Duffy08}, whereas we used the relation from Neto et al. (2007) for the other cosmologies. For a step-by-step description of how to use the mass history models (analytic and semi-analytic) that were presented in Sections 2 and 4.4, respectively, see Appendix~\ref{MAH_Description}.

\subsubsection{Mass definition dependence}\label{Massdef}

So far our calculations have been based on a halo mass defined as the mass of all matter within the radius $r_{200}$ at which the mean internal density $\langle\rho\rangle(<r_{200})$ is a factor of $\Delta=200$ times the critical density of the universe, $\rho_{\rm crit}$ (from now on we denote this halo mass by $M_{\rm{200}}$). In the literature a number of values have been used for $\Delta$. Some authors opt to use $\Delta=200$ (e.g., \citealt{Jenkins}) or $\Delta=200\Omega_{\rm{m}}(z)$ (e.g. NFW), while others (e.g., \citealt{Bullock}) choose $\Delta=\Delta_{\rm{vir}}$ according to the spherical  virialization criterion of \citet{BryanNorman}. These definitions can lead to sizeable differences in $c$ for a given halo and, as discussed, the differences are also cosmology-dependent. 

In this section we study how the structural properties and mass accretion histories depend on the adopted mass definition. We analyse halo mass histories of relaxed halos using three different halo mass definitions. First, we use $M_{\rm{200}}$. Secondly, we use $M_{\rm{mean}}$, which is the mass within the radius $r_{\rm{mean}}$ for which the mean internal density is 200 times the mean background density. Finally, $M_{\rm{vir}}$ is the mass within the radius $r_{\rm{vir}}$ for which the mean internal density is $\Delta_{\rm{vir}}$ times the critical density as determined by \citet{BryanNorman}. Note that halo masses and radii are determined using a spherical overdensity routine within the SUBFIND algorithm (\citealt{Springel}) centred on the main subhalo of the FoF halos (\citealt{Davis}). We perform all calculations for the three different halo definitions, taking the halo centre to be the location of the particle in the FoF group for which the gravitational potential is minimum.

Eq.~(\ref{rho-crit}) shows that the formation redshift is directly proportional to the mean density within the scale radius ($(1+z_{-2})^{3}\propto \langle\rho\rangle(<r_{-2})$), but the constant of proportionality depends on the mass definition that is adopted. Therefore, a change in the mass definition changes the formation time as
\begin{equation}\label{massdef}
\frac{(1+z_{f})_{\Delta_{1}}}{(1+z_{f})_{\Delta_{2}}}\approx \left(\frac{\langle\rho\rangle(<r_{-2})_{\Delta_{1}}}{\langle\rho\rangle(<r_{-2})_{\Delta_{2}}}\right)^{1/3},
\end{equation}

\noindent where $\Delta_{1,2}$ refers to different overdensity criteria. That is, from eq. (\ref{rho-2}), $\langle\rho\rangle(<r_{-2})$ changes according to the mass definition as $\langle\rho\rangle(<r_{-2})_{\Delta}=\Delta\times\rho_{\rm{crit},0}c_{\Delta}^{3}Y(1)/Y(c_{\Delta})$. Then ${\langle\rho\rangle(<r_{-2})_{\Delta_{1}=200}}$ refers to the mean internal density within $r_{-2}$, obtained by defining the mean internal density at the virial radius to be $200\rho_{\rm{crit},0}$.  If we consider $\Delta_{1}=200$ and $\Delta_{2}=\rm{mean}=200\Omega_{m}$, and that there is a factor of $0.55$ difference between the concentrations $c_{\rm{200}}$ and $c_{\rm{mean}}$ for a $10^{12}\rm{M}_{\sun}$ halo, then we obtain (using eq. \ref{massdef}) the relation ${(1+z_{\rm{f}})_{200}\approx(0.255\Omega_{\rm{m}}^{-1})^{1/3}(1+z_{\rm{f}})_{\rm{mean}}}$, where we have used the fact that ${\langle\rho\rangle(<r_{-2})_{\rm{mean}}\sim(\Omega_{\rm{m}}/0.255)
\langle\rho\rangle(<r_{-2})_{200}}$\footnote{In this last step we used the approximation that $\langle\rho\rangle(<r_{-2})_{\Delta}=\Delta\times c^{3}Y(1)/Y(c)\rho_{\rm{crit},0}\approx \Delta\times 0.643c^{2.28}\rho_{\rm{crit},0}$}. This implies that the change in the halo mass history due to different halo mass definitions is

\begin{eqnarray}\nonumber 
\log_{10}\frac{M(z)_{200}}{M(z)_{\rm{mean}}} &=& \log_{10}(1+z)^{\alpha_{200}-\alpha_{\rm{mean}}}\\
&& +\log_{10}e^{(\beta_{200}-\beta_{\rm{mean}})z}\\\label{anterior}
&\approx & 0.956\alpha_{200}\log_{10}(1+z)\\\nonumber
&& +[1-(0.255/\Omega_{\rm{m}})^{1/3}]\log_{10}(e)\beta_{200}z\\\label{DeltaMassdef}
&\approx & 0.0543 z,
\end{eqnarray}

\noindent where in step (\ref{anterior}) we replaced $\alpha_{200}-\alpha_{\rm{mean}}\approx 0.956\alpha_{200}$, which is valid for a $10^{12}\rm{M}_{\sun}$ halo, and $\alpha_{200}=0.2501$ and $\beta_{200}=-0.8147$.

\begin{figure} 
\begin{center}
\includegraphics[angle=0,width=0.48\textwidth]{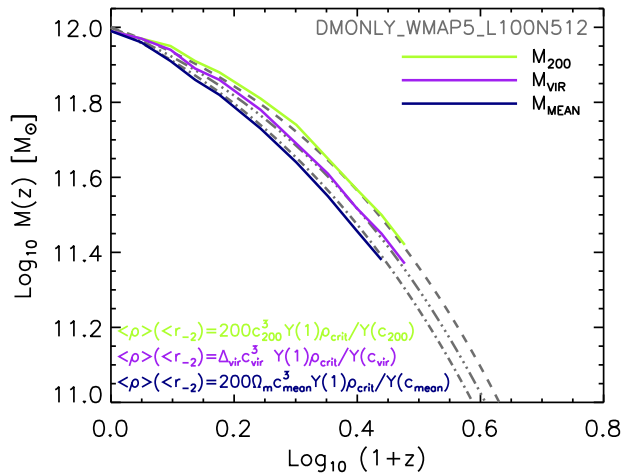}%
\caption{Mass history of a $10^{12}\rm{M}_{\sun}$ halo as a function of redshift. The different coloured lines show the change in the mass history when different halo mass definitions are used. The green line shows the mass history of a halo of $M_{\rm{200}}=10^{12}\rm{M}_{\sun}$ at $z=0$, whereas the dark blue (purple) line shows the mass history of a halo of $M_{\rm{mean}}=10^{12}\rm{M}_{\sun}$ ($M_{\rm{vir}}=10^{12}\rm{M}_{\sun}$) at $z=0$. The dashed lines show the mass history predicted by eqs.~(\ref{final_set1}), (\ref{final_set2}) and (\ref{final_set3}). The difference lies in the formation redshift definition which is affected by the change in the mean inner density (see eq.~\ref{rho-crit}). The value of $\langle\rho\rangle(<r_{-2})$ changes with the value of $\Delta$ we used in the definition of halo mass. The $c-M$ relation correspond to the mass definition under consideration.}
\label{Massdefinition}
\end{center}
\end{figure}

The difference in mass history given by eq. (\ref{DeltaMassdef}) can be seen in Fig. \ref{Massdefinition}, which shows how the halo mass history is affected by the halo mass definition. The green line in Fig. \ref{Massdefinition} shows the mass history assuming the $M_{\rm{200}}$ mass definition. The purple line shows the $M_{\rm{vir}}$ definition, and the dark blue line shows the $M_{\rm{mean}}$ definition. The different dashed lines correspond to the mass histories $M(z)=M(z=0)(1+z)^{\alpha}e^{\beta z}$, where the difference lies in the mass definition that changes the mean inner density and the concentration-mass relation (for a relaxed halo sample). Duffy et al. (2008) studied how the halo mass definition changes the concentration-mass relation, and provided the parameters of the different $c-M$ relations. They found that the concentration of a relaxed $M_{\rm{mean}}$ halo is $80\%$ larger than the concentration of a relaxed $M_{\rm{200}}$ halo. We adopt those fits in our calculations of the $M(z)$ estimate and conclude that, as long as we use a concentration$-$mass relation that is consistent with the adopted halo mass definition, the expressions (\ref{final_set1}), (\ref{final_set2}) and (\ref{final_set3}) accurately reproduce the halo mass history.

The analytic estimate given by eq.~(\ref{DeltaMassdef}) predicts that the difference in mass history due to the change in the mass definition ($M_{\rm{200}}$ versus $M_{\rm{mean}}$) is $\Delta\log_{10} M(z)\approx 0.0543 z$. This can be seen in Fig. \ref{Massdefinition}, where ${\Delta\log_{10} M(z)=0.1086}$ at $z=2$ (${M_{\rm{mean}}(z=2)=10^{11.3035}\rm{M}_{\sun}}$ and ${M_{\rm{200}}(z=2)=10^{11.4117}\rm{M}_{\sun}}$).

\section{Comparison between semi-analytic and analytic models}\label{compare_models}

In this section we compare the semi-analytic model derived in Section \ref{MAH_semianalytic}, with the analytic model for halo mass history derived in Paper I. Note that while the semi-analytic model is obtained through fits to simulations, the analytic model was based on the extended Press-Schechter (EPS) theory without calibration against simulations.

\begin{figure} 
\begin{center}
\includegraphics[angle=0,width=0.48\textwidth]{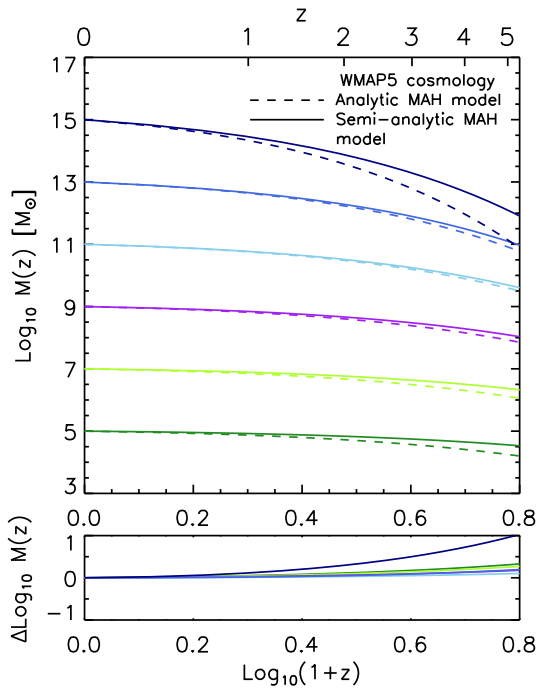}%
\caption{The top panel shows a comparison between the semi-analytic (solid lines) and the analytic model (dashed lines) for halo mass history. The bottom panel shows the residuals between the models. The different coloured lines correspond to the models for various halo masses in range $\log_{10}[M_{0}/\rm{M}_{\sun}]=5-15$ in steps of 2.}
\label{Comparison_models}
\end{center}
\end{figure}

Fig.~\ref{Comparison_models} shows a comparison between the models for various halo masses ($\log_{10}[M_{0}/\rm{M}_{\sun}]=5-15$). As can be seen from the figure, the models mostly agree on the mass histories of halos with final masses between $10^{10}$ and $10^{14}\rm{M}_{\sun}$. However, there are a few factors of difference in the mass histories of larger and smaller halos, and the difference increases towards high redshift. We find that in the analytic model the halo mass decreases quite abruptly at high redshift for halos with final masses $>10^{14}\rm{M}_{\sun}$. For instance, there is a factor of 9 difference at $z=5$ between the models for a $10^{15}\rm{M}_{\sun}$ halo. This difference is probably due to the progenitor definition. In the analytic model, the progenitor is defined as the halo with mass a factor of $q$ lower ($q\sim 4$ for $M_{0}>10^{14}\rm{M}_{\sun}$) at redshift $z_{f}$ ($z_{f}\sim 0.9$), whereas in the semi-analytic model, the progenitor is the halo that contains most of the 25 most bound dark matter particles in the following snapshot.

Fig.~\ref{Comparison_models} also shows that the semi-analytic model overpredicts the mass histories of halos with final masses $<10^{9}\rm{M}_{\sun}$. This is expected because the parameters $\alpha$ and $\beta$ in the semi-analytic model depend on the concentration-mass relation adopted. In this case we are using Duffy et al. (2008) relation, which is calibrated in the mass range $10^{10}-10^{14}\rm{M}_{\sun}$, for lower masses the concentration-mass relation deviates from a simple power law (\citealt{Ludlow14}).

In Paper I we developed an analytic model of halo mass history, based on EPS theory, that only depends on the power spectrum of the primordial density perturbations. We found very good agreement between the halo mass histories predicted by our analytic model and published fits to simulation results (\citealt{van,McBride,vandenBosch14}). In this work we have developed a semi-analytic model that uses the functional form for the mass history motivated by EPS theory, and linked the mass history to halo structure through empirical relations obtained from simulations. 

We now combine these two models (semi-analytic and analytic) to establish the physical link between a halo concentration and the linear {\it{rms}} fluctuation of the primordial density field. From Fig.~\ref{Comparison_models} we have found that there are a few factors of difference between the models. We now focus on the mass range $10^{11}-10^{15}\rm{M}_{\sun}$, where the factor of difference is less than 1.5. We set the mass history curve to be the same in the two models, that is 

\begin{equation}
M(z)_{\rm{Analytic}} = M(z)_{\rm{Semi-analytic}}
\end{equation}

\noindent for all redshifts. We then evaluate this equality at redshift 1 and obtain,

\begin{eqnarray}\label{sigma_c}
f(M_{0})\left(0.92\frac{dD}{dz}|_{z=0}-0.3\right) &=& \alpha_{\rm{S}}(c)\ln(2)+\beta_{\rm{S}}(c).
\end{eqnarray}

\begin{figure} 
\begin{center}
\includegraphics[angle=0,width=0.4\textwidth]{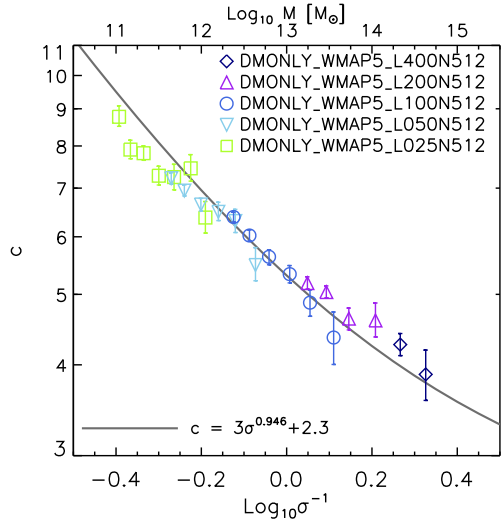}%
\caption{Comparison between the $\sigma-c$ relation at $z=0$ predicted by the combination of the mass history models (solid line), and the simulation outputs (coloured symbols).}
\label{sigma_c_fig}
\end{center}
\end{figure}

\noindent In this last equation, $\alpha_{\rm{S}}(c)$ and $\beta_{\rm{S}}(c)$ are given by eqs. (\ref{alpha3}) and (\ref{beta}), respectively, and depend on concentration, $D$ is the linear growth factor and $f(M_{0})$ depends on the rms of the primordial density field, $\sigma$. We approximate various terms in eq. (\ref{sigma_c}), including $f(M_{0})\sim 1.155 (\sigma(M_{0})^{2})^{0.277}$ and $Y(1)/Y(c)\sim 0.643 c^{-0.71}$, and obtain

\begin{equation}\label{c_sigma_eq}
c = 3\sigma^{0.946}+2.3,
\end{equation}

\noindent which is suitable for a WMAP5 cosmology. Note that eq.~(\ref{c_sigma_eq}) is not a fit to any simulation data, it has been derived from eq.~(\ref{sigma_c}). Fig.~\ref{sigma_c_fig} shows the $c-\sigma$ relation at $z=0$. In this figure we compare the predicted relation (solid line), as given by eq.~(\ref{c_sigma_eq}) (obtained by equaling the analytic and semi-analytic models), with the simulation outputs (coloured symbols). The different symbols correspond to the median values of the relaxed sample of halos and the error bars to $1\sigma$ confidence limits. The good agreement between the analytic prediction and the simulation outputs clearly shows that the halo mass accretion history is the physical connection in the $c-\sigma$ correlation.

\section{Summary and conclusion}\label{SummaryAndConclussion}

In this work we have demonstrated that there is an intrinsic relation between halo assembly history and inner halo structure, and that the mass history is the physical connection between the inner halo structure and the power spectrum of initial density fluctuations.

We examined the density profiles and mass growth histories of a large sample of halos and their progenitors within the OWLS simulations. We separated our halo sample into a `relaxed' sample, and a `complete' sample that includes both relaxed and unrelaxed halos. We confirmed the finding of \citet{Ludlow} that for relaxed halos the mean enclosed density within the NFW scale radius ($r_{-2}$), $\langle\rho\rangle(<r_{-2})$,  is directly proportional to the critical density of the Universe at the formation redshift, $z_{-2}$, defined as the time at which the mass of the main progenitor equals the mass enclosed within the scale radius at $z=0$,

\begin{equation}\label{conclu1}
\langle\rho\rangle(<r_{-2}) = 900\rho_{\rm{crit}}(z_{-2}).
\end{equation}

\noindent In the above relation, the value, $900\pm50$, is obtained through fits to the simulation data (suitable for WMAP5 cosmology). We showed in Section \ref{cosmo_sec} that expression (\ref{conclu1}) provides a straightforward relation between formation time and concentration,

\begin{equation}
(1+z_{-2}) = [(200/900)c^{3}Y(1)/Y(c)-\Omega_{\Lambda}]^{1/3}\Omega_{\rm{m}}^{-1/3},
\end{equation}

\noindent where $Y(c)=\ln(1+c)-c/(1+c)$. The overall trend of decreasing formation redshift with decreasing concentration is evident for the main branches of the descendant halos. This implies that halos which assemble earlier have a higher concentration, because the density of the Universe at the formation time was larger. To relate the formation time to halo mass, we used the following concentration mass relation for relaxed halos

\begin{equation}
c = 6.67(M/2\times 10^{12}h^{-1}\rm{M}_{\sun})^{-0.092},
\end{equation}

\noindent obtained by \citet{Duffy08} by fitting to the simulation data and suitable for the WMAP5 cosmology. We found that, on average, halo concentrations differ by a factor of $1.16$ between the relaxed and complete samples. The lower individual concentrations of unrelaxed halos (due to spurious subhalos or ongoing mergers that do not result in an accurate fit for an NFW density profile) produce incorrect enclosed halo masses and therefore lower formation times (by a factor of $1.1$). However, on average, the formation time-concentration relation does not change, thus indicating that the halo mass history is not affected by the fact that a halo is out of equilibrium at a particular redshift. We used these findings to show that scatter in the halo mass history leads to scatter in the formation time ($\delta z_{-2}$), and hence to scatter in the concentration$-$mass relation ($\delta c$). 

We found that formation time decreases with increasing mass (at a non-linear rate). This means that high-mass halos are still accreting mass rapidly in the present epoch, while low-mass halos typically accreted their mass early. Thus, the formation time$-$concentration relation provides the physical link between the halo mass history and internal structure. This result led us to provide a semi-analytic model for the halo mass history, which uses a direct, analytic correlation between the parameters $\alpha$ and $\beta$ in the mass history (eq.~\ref{final_set1}) and concentration, 

\begin{eqnarray}\label{final_set1}
M(z) &=& M_{0}(1+z)^{\alpha}e^{\beta z},\\\label{final_set2}
\alpha &=& [\ln(Y(1)/Y(c))-\beta z_{-2}]/\ln(1+z_{-2}),\\\label{final_set3}
\beta &=& -3/(1+z_{-2}),
\end{eqnarray}

\noindent where we obtained the constant value, $-3.01\pm 0.08$, in the last relation (eq. \ref{final_set3}) by fitting the halo mass history model (eq. \ref{final_set1}) to the simulation data. Then we obtained a semi-analytic model for the mass accretion history, that adopts the functional form, eq. (\ref{final_set1}), and the parameters $\alpha$ and $\beta$ are now given by analytic relations that include numbers obtained from fits to simulation results. It is important to note that the semi-analytic model was derived assuming that the density profile of all halos is NFW. Interestingly, we found that the semi-analytic model describes with high-accuracy the mass histories of both relaxed and unrelaxed halo sample, even though the NFW profile is not a good fit to the density profile of unrelaxed halos.

We investigated how cosmology affects the semi-analytic model. We found that as long as a suitable concentration-mass relation and the value for the best-fitting parameter in the ${\langle\rho\rangle(<r_{-2})-\rho_{\rm{crit}}(z_{-2})}$ relation are assumed for the cosmology being considered, the semi-analytic models describe the mass histories with high accuracy. In addition, we investigated how different mass definitions change the halo mass histories and we found that as long as we use a concentration-mass relation that is consistent with the adopted halo mass definition, the semi-analytic model accurately reproduces the halo mass history.

In Paper I, we presented an analytic model for the halo mass history, based on extended Press-Schechter theory and not calibrated against simulations data. We compared the analytic model of Paper I with the semi-analytic model presented here and found very good agreement in the mass range $10^{9}-10^{14}\rm{M}_{\sun}$. However, we found that the analytic model predicts larger masses at high redshift for halos with final masses $>10^{14}\rm{M}_{\sun}$, whereas the semi-analytic model overpredicts the mass history of low-mass halos (halos with final masses $<10^{9}\rm{M}_{\sun}$). This is expected because the semi-analytic model depends on the adopted concentration-mass relation, which deviates from the assumed power law at low masses. The reader may find a step-by-step guide on how to implement the semi-analytic model in Appendix \ref{MAH_Description}, as well as numerical routines online\footnote{Available at \href{https://bitbucket.org/astroduff/commah}{\it{https://bitbucket.org/astroduff/commah}}.}.

Interestingly, by combining these two models (semi-analytic and analytic) we established the physical link between a halo concentration and the initial density perturbation field, which explains the correlation between concentration and {\it{rms}} fluctuation of the primordial density field, $\sigma$ (Fig.~\ref{sigma_c_fig}). 

Finally, by differentiating eq.~(\ref{final_set1}) we obtained the dark matter accretion rate,

\begin{eqnarray}\nonumber
\frac{dM}{dt} &=& 71.6 \rm{M}_{\sun}\rm{yr}^{-1} M_{12}h_{0.7}\\
&& \times[-\alpha-\beta(1+z)][\Omega_{m}(1+z)^{3}+\Omega_{\Lambda}]^{1/2}.
\end{eqnarray}

\noindent As the change in the $\alpha$ and $\beta$ parameters over halo masses is not significant when calculating accretion rates, we provided a {\it{mean}} accretion rate, obtained by averaging the parameters $\alpha$ and $\beta$ over halo mass,  

\begin{eqnarray}\nonumber
\langle\frac{dM}{dt}\rangle &=& 71.6 \rm{M}_{\sun}\rm{yr}^{-1} M_{12}h_{0.7}\\
&& \times[-0.24+0.75(1+z)][\Omega_{\rm{m}}(1+z)^{3}+\Omega_{\Lambda}]^{1/2}.
\end{eqnarray}

\noindent We then concluded that in order to predict halo mass growth, the concentration-mass relation from a relaxed sample should be used.

Putting the pieces together, we addressed the question of how the structure of halos depends on the primordial density perturbation field. We found that concentration is the link between the halo mass profile and the halo mass history (and that one can be determined from the other). We also found that the `shape' of the halo mass history is given by the linear growth factor and linear power spectrum of density fluctuations. Therefore, we concluded that halo concentrations are directly connected to the initial density perturbation field. 

In a forthcoming paper (Paper III) we will combine the analytic model presented in Paper I and semi-analytic model presented here to predict the concentration-mass relation. We will investigate its evolution. We will show that extrapolations to low masses of power-law fits to simulation results are highly inadequate, and will investigate whether linear ${\langle\rho\rangle(<r_{-2})-\rho_{\rm{crit}}(z_{-2})}$ relation holds at redshifts other than 0.

\section*{Acknowledgements}

We are grateful to the referee Aaron Ludlow for fruitful comments that substantially improved the original manuscript. CAC acknowledges the support of the 2013 John Hodgson Scholarship and the hospitality of Leiden Observatory. JSBW is supported by an Australian Research Council Laureate Fellowship. JS acknowledges support by the European Research Council under the European  Union's Seventh Framework Programme (FP7/2007-2013)/ERC Grant agreement 278594-GasAroundGalaxies. We are grateful to the OWLS team for their help with the simulations. 

\bibliography{biblio}

\begin{thebibliography}{}

\bibitem[\protect\citeauthoryear{{Bryan} \& {Norman}}{{Bryan} \&
  {Norman}}{1998}]{BryanNorman}
{Bryan} G.~L.,  {Norman} M.~L.,  1998, \apj, 495, 80

\bibitem[\protect\citeauthoryear{{Bullock}, {Kolatt}, {Sigad}, {Somerville},
  {Kravtsov}, {Klypin}, {Primack} \& {Dekel}}{{Bullock} et~al.}{2001}]{Bullock}
{Bullock} J.~S.,  {Kolatt} T.~S.,  {Sigad} Y.,  {Somerville} R.~S.,  {Kravtsov}
  A.~V.,  {Klypin} A.~A.,  {Primack} J.~R.,    {Dekel} A.,  2001, \mnras, 321,
  559

\bibitem[\protect\citeauthoryear{{Correa}, {Wyithe}, {Schaye} \&
  {Duffy}}{{Correa} et~al.}{2015a}]{PaperI}
{Correa} C.~A.,  {Wyithe} J.~S.~B.,  {Schaye} J.,    {Duffy} A.~R.,  2015a,
  \mnras, 450, 1514 (Paper I)

\bibitem[\protect\citeauthoryear{{Correa}, {Wyithe}, {Schaye} \&
  {Duffy}}{{Correa} et~al.}{2015c}]{PaperIII}
{Correa} C.~A.,  {Wyithe} J.~S.~B.,  {Schaye} J.,    {Duffy} A.~R.,  2015c,
  ArXiv e-prints:1502.00391 (Paper III)

\bibitem[\protect\citeauthoryear{{Dalal}, {Lithwick} \& {Kuhlen}}{{Dalal}
  et~al.}{2010}]{Dalal}
{Dalal} N.,  {Lithwick} Y.,    {Kuhlen} M.,  2010, ArXiv e-prints:1010.2539

\bibitem[\protect\citeauthoryear{{Dalla Vecchia} \& {Schaye}}{{Dalla Vecchia}
  \& {Schaye}}{2008}]{DallaVecchia}
{Dalla Vecchia} C.,  {Schaye} J.,  2008, \mnras, 387, 1431

\bibitem[\protect\citeauthoryear{{Davis}, {Efstathiou}, {Frenk} \&
  {White}}{{Davis} et~al.}{1985}]{Davis}
{Davis} M.,  {Efstathiou} G.,  {Frenk} C.~S.,    {White} S.~D.~M.,  1985, \apj,
  292, 371

\bibitem[\protect\citeauthoryear{{Dekel}, {Devor} \& {Hetzroni}}{{Dekel}
  et~al.}{2003}]{Dekel03}
{Dekel} A.,  {Devor} J.,    {Hetzroni} G.,  2003, \mnras, 341, 326

\bibitem[\protect\citeauthoryear{{Dekel} \& {Krumholz}}{{Dekel} \&
  {Krumholz}}{2013}]{Dekel13}
{Dekel} A.,  {Krumholz} M.~R.,  2013, \mnras, 432, 455

\bibitem[\protect\citeauthoryear{{Diemer} \& {Kravtsov}}{{Diemer} \&
  {Kravtsov}}{2014}]{Diemer14a}
{Diemer} B.,  {Kravtsov} A.~V.,  2014, \apj, 789, 1

\bibitem[\protect\citeauthoryear{{Diemer} \& {Kravtsov}}{{Diemer} \&
  {Kravtsov}}{2015}]{Diemer14}
{Diemer} B.,  {Kravtsov} A.~V.,  2015, \apj, 799, 108

\bibitem[\protect\citeauthoryear{{Diemer}, {More} \& {Kravtsov}}{{Diemer}
  et~al.}{2013}]{Diemer}
{Diemer} B.,  {More} S.,    {Kravtsov} A.~V.,  2013, \apj, 766, 25

\bibitem[\protect\citeauthoryear{{Dolag}, {Borgani}, {Murante} \&
  {Springel}}{{Dolag} et~al.}{2009}]{Dolag}
{Dolag} K.,  {Borgani} S.,  {Murante} G.,    {Springel} V.,  2009, \mnras, 399,
  497

\bibitem[\protect\citeauthoryear{{Duffy}, {Schaye}, {Kay} \& {Dalla
  Vecchia}}{{Duffy} et~al.}{2008}]{Duffy08}
{Duffy} A.~R.,  {Schaye} J.,  {Kay} S.~T.,    {Dalla Vecchia} C.,  2008,
  \mnras, 390, L64

\bibitem[\protect\citeauthoryear{{Dutton} \& {Macci{\`o}}}{{Dutton} \&
  {Macci{\`o}}}{2014}]{Dutton14}
{Dutton} A.~A.,  {Macci{\`o}} A.~V.,  2014, \mnras, 441, 3359

\bibitem[\protect\citeauthoryear{{Einasto}}{{Einasto}}{1965}]{Einasto}
{Einasto} J.,  1965, Trudy Astrofizicheskogo Instituta Alma-Ata, 5, 87

\bibitem[\protect\citeauthoryear{{Eke}, {Navarro} \& {Steinmetz}}{{Eke}
  et~al.}{2001}]{Eke}
{Eke} V.~R.,  {Navarro} J.~F.,    {Steinmetz} M.,  2001, \apj, 554, 114

\bibitem[\protect\citeauthoryear{{Fakhouri}, {Ma} \&
  {Boylan-Kolchin}}{{Fakhouri} et~al.}{2010}]{Fakhouri}
{Fakhouri} O.,  {Ma} C.-P.,    {Boylan-Kolchin} M.,  2010, \mnras, 406, 2267

\bibitem[\protect\citeauthoryear{{Giocoli}, {Tormen} \& {Sheth}}{{Giocoli}
  et~al.}{2012}]{Giocoli}
{Giocoli} C.,  {Tormen} G.,    {Sheth} R.~K.,  2012, \mnras, 422, 185

\bibitem[\protect\citeauthoryear{{Hayashi} \& {White}}{{Hayashi} \&
  {White}}{2008}]{Hayashi}
{Hayashi} E.,  {White} S.~D.~M.,  2008, \mnras, 388, 2

\bibitem[\protect\citeauthoryear{{Hinshaw}, {Larson}, {Komatsu}, {Spergel},
  {Bennett}, {Dunkley}, {Nolta}, {Halpern}, {Hill}, {Odegard} \& et
  al.}{{Hinshaw} et~al.}{2013}]{Hinshaw}
{Hinshaw} G.,  {Larson} D.,  {Komatsu} E.,  {Spergel} D.~N.,  {Bennett} C.~L.,
  {Dunkley} J.,  {Nolta} M.~R.,  {Halpern} M.,  {Hill} R.~S.,  {Odegard} N.,
  et al. 2013, \apjs, 208, 19

\bibitem[\protect\citeauthoryear{{Huss}, {Jain} \& {Steinmetz}}{{Huss}
  et~al.}{1999}]{Huss}
{Huss} A.,  {Jain} B.,    {Steinmetz} M.,  1999, \apj, 517, 64

\bibitem[\protect\citeauthoryear{{Jenkins}, {Frenk}, {White}, {Colberg},
  {Cole}, {Evrard}, {Couchman} \& {Yoshida}}{{Jenkins} et~al.}{2001}]{Jenkins}
{Jenkins} A.,  {Frenk} C.~S.,  {White} S.~D.~M.,  {Colberg} J.~M.,  {Cole} S.,
  {Evrard} A.~E.,  {Couchman} H.~M.~P.,    {Yoshida} N.,  2001, \mnras, 321,
  372

\bibitem[\protect\citeauthoryear{{Klypin}, {Trujillo-Gomez} \&
  {Primack}}{{Klypin} et~al.}{2011}]{Klypin}
{Klypin} A.~A.,  {Trujillo-Gomez} S.,    {Primack} J.,  2011, \apj, 740, 102

\bibitem[\protect\citeauthoryear{{Komatsu}, {Dunkley}, {Nolta}, {Bennett},
  {Gold}, {Hinshaw}, {Jarosik}, {Larson}, {Limon}, {Page} \& et al.}{{Komatsu}
  et~al.}{2009}]{Komatsu}
{Komatsu} E.,  {Dunkley} J.,  {Nolta} M.~R.,  {Bennett} C.~L.,  {Gold} B.,
  {Hinshaw} G.,  {Jarosik} N.,  {Larson} D.,  {Limon} M.,  {Page} L.,    et al.
  2009, \apjs, 180, 330

\bibitem[\protect\citeauthoryear{{Kuhlen}, {Strigari}, {Zentner}, {Bullock} \&
  {Primack}}{{Kuhlen} et~al.}{2005}]{Kuhlen}
{Kuhlen} M.,  {Strigari} L.~E.,  {Zentner} A.~R.,  {Bullock} J.~S.,
  {Primack} J.~R.,  2005, \mnras, 357, 387

\bibitem[\protect\citeauthoryear{{Lacerna} \& {Padilla}}{{Lacerna} \&
  {Padilla}}{2011}]{Lacerna}
{Lacerna} I.,  {Padilla} N.,  2011, \mnras, 412, 1283

\bibitem[\protect\citeauthoryear{{Lu}, {Mo}, {Katz} \& {Weinberg}}{{Lu}
  et~al.}{2006}]{Lu2006}
{Lu} Y.,  {Mo} H.~J.,  {Katz} N.,    {Weinberg} M.~D.,  2006, \mnras, 368, 1931

\bibitem[\protect\citeauthoryear{{Ludlow}, {Navarro}, {Angulo},
  {Boylan-Kolchin}, {Springel}, {Frenk} \& {White}}{{Ludlow}
  et~al.}{2014}]{Ludlow14}
{Ludlow} A.~D.,  {Navarro} J.~F.,  {Angulo} R.~E.,  {Boylan-Kolchin} M.,
  {Springel} V.,  {Frenk} C.,    {White} S.~D.~M.,  2014, \mnras, 441, 378

\bibitem[\protect\citeauthoryear{{Ludlow}, {Navarro}, {Boylan-Kolchin}, {Bett},
  {Angulo}, {Li}, {White}, {Frenk} \& {Springel}}{{Ludlow}
  et~al.}{2013}]{Ludlow}
{Ludlow} A.~D.,  {Navarro} J.~F.,  {Boylan-Kolchin} M.,  {Bett} P.~E.,
  {Angulo} R.~E.,  {Li} M.,  {White} S.~D.~M.,  {Frenk} C.,    {Springel} V.,
  2013, \mnras, 432, 1103

\bibitem[\protect\citeauthoryear{{Ludlow}, {Navarro}, {Li}, {Angulo},
  {Boylan-Kolchin} \& {Bett}}{{Ludlow} et~al.}{2012}]{Ludlow12}
{Ludlow} A.~D.,  {Navarro} J.~F.,  {Li} M.,  {Angulo} R.~E.,  {Boylan-Kolchin}
  M.,    {Bett} P.~E.,  2012, \mnras, 427, 1322

\bibitem[\protect\citeauthoryear{{Ludlow}, {Navarro}, {Springel},
  {Vogelsberger}, {Wang}, {White}, {Jenkins} \& {Frenk}}{{Ludlow}
  et~al.}{2010}]{Ludlow10}
{Ludlow} A.~D.,  {Navarro} J.~F.,  {Springel} V.,  {Vogelsberger} M.,  {Wang}
  J.,  {White} S.~D.~M.,  {Jenkins} A.,    {Frenk} C.~S.,  2010, \mnras, 406,
  137

\bibitem[\protect\citeauthoryear{{Macci{\`o}}, {Dutton} \& {van den
  Bosch}}{{Macci{\`o}} et~al.}{2008}]{Maccio08}
{Macci{\`o}} A.~V.,  {Dutton} A.~A.,    {van den Bosch} F.~C.,  2008, \mnras,
  391, 1940

\bibitem[\protect\citeauthoryear{{Macci{\`o}}, {Dutton}, {van den Bosch},
  {Moore}, {Potter} \& {Stadel}}{{Macci{\`o}} et~al.}{2007}]{Maccio07}
{Macci{\`o}} A.~V.,  {Dutton} A.~A.,  {van den Bosch} F.~C.,  {Moore} B.,
  {Potter} D.,    {Stadel} J.,  2007, \mnras, 378, 55

\bibitem[\protect\citeauthoryear{{Manrique}, {Raig}, {Salvador-Sol{\'e}},
  {Sanchis} \& {Solanes}}{{Manrique} et~al.}{2003}]{Manrique}
{Manrique} A.,  {Raig} A.,  {Salvador-Sol{\'e}} E.,  {Sanchis} T.,    {Solanes}
  J.~M.,  2003, \apj, 593, 26

\bibitem[\protect\citeauthoryear{{McBride}, {Fakhouri} \& {Ma}}{{McBride}
  et~al.}{2009}]{McBride}
{McBride} J.,  {Fakhouri} O.,    {Ma} C.-P.,  2009, \mnras, 398, 1858

\bibitem[\protect\citeauthoryear{{Navarro}, {Frenk} \& {White}}{{Navarro}
  et~al.}{1996}]{NFW96}
{Navarro} J.~F.,  {Frenk} C.~S.,    {White} S.~D.~M.,  1996, \apj, 462, 563 (NFW)

\bibitem[\protect\citeauthoryear{{Navarro}, {Frenk} \& {White}}{{Navarro}
  et~al.}{1997}]{NFW97}
{Navarro} J.~F.,  {Frenk} C.~S.,    {White} S.~D.~M.,  1997, \apj, 490, 493

\bibitem[\protect\citeauthoryear{{Navarro}, {Hayashi}, {Power}, {Jenkins},
  {Frenk}, {White}, {Springel}, {Stadel} \& {Quinn}}{{Navarro}
  et~al.}{2004}]{Navarro04}
{Navarro} J.~F.,  {Hayashi} E.,  {Power} C.,  {Jenkins} A.~R.,  {Frenk} C.~S.,
  {White} S.~D.~M.,  {Springel} V.,  {Stadel} J.,    {Quinn} T.~R.,  2004,
  \mnras, 349, 1039

\bibitem[\protect\citeauthoryear{{Navarro}, {Ludlow}, {Springel}, {Wang},
  {Vogelsberger}, {White}, {Jenkins}, {Frenk} \& {Helmi}}{{Navarro}
  et~al.}{2010}]{Navarro10}
{Navarro} J.~F.,  {Ludlow} A.,  {Springel} V.,  {Wang} J.,  {Vogelsberger} M.,
  {White} S.~D.~M.,  {Jenkins} A.,  {Frenk} C.~S.,    {Helmi} A.,  2010,
  \mnras, 402, 21

\bibitem[\protect\citeauthoryear{{Neistein} \& {Dekel}}{{Neistein} \&
  {Dekel}}{2008}]{Neistein08}
{Neistein} E.,  {Dekel} A.,  2008, \mnras, 388, 1792

\bibitem[\protect\citeauthoryear{{Neistein}, {van den Bosch} \&
  {Dekel}}{{Neistein} et~al.}{2006}]{Neistein}
{Neistein} E.,  {van den Bosch} F.~C.,    {Dekel} A.,  2006, \mnras, 372, 933

\bibitem[\protect\citeauthoryear{{Neto}, {Gao}, {Bett}, {Cole}, {Navarro},
  {Frenk}, {White}, {Springel} \& {Jenkins}}{{Neto} et~al.}{2007}]{Neto}
{Neto} A.~F.,  {Gao} L.,  {Bett} P.,  {Cole} S.,  {Navarro} J.~F.,  {Frenk}
  C.~S.,  {White} S.~D.~M.,  {Springel} V.,    {Jenkins} A.,  2007, \mnras,
  381, 1450

\bibitem[\protect\citeauthoryear{{Planck Collaboration}, {Ade}, {Aghanim},
  {Armitage-Caplan}, {Arnaud}, {Ashdown}, {Atrio-Barandela}, {Aumont},
  {Baccigalupi}, {Banday} \& et al.}{{Planck Collaboration}
  et~al.}{2014}]{Planck}
{Planck Collaboration} {Ade} P.~A.~R.,  {Aghanim} N.,  {Armitage-Caplan} C.,
  {Arnaud} M.,  {Ashdown} M.,  {Atrio-Barandela} F.,  {Aumont} J.,
  {Baccigalupi} C.,  {Banday} A.~J.,    et al. 2014, \aap, 571, A16

\bibitem[\protect\citeauthoryear{{Prada}, {Klypin}, {Cuesta}, {Betancort-Rijo}
  \& {Primack}}{{Prada} et~al.}{2012}]{Prada}
{Prada} F.,  {Klypin} A.~A.,  {Cuesta} A.~J.,  {Betancort-Rijo} J.~E.,
  {Primack} J.,  2012, \mnras, 423, 3018

\bibitem[\protect\citeauthoryear{{Salvador-Sol{\'e}}, {Vi{\~n}as}, {Manrique}
  \& {Serra}}{{Salvador-Sol{\'e}} et~al.}{2012}]{Salvador}
{Salvador-Sol{\'e}} E.,  {Vi{\~n}as} J.,  {Manrique} A.,    {Serra} S.,  2012,
  \mnras, 423, 2190

\bibitem[\protect\citeauthoryear{{Schaye} \& {Dalla Vecchia}}{{Schaye} \&
  {Dalla Vecchia}}{2008}]{Schaye08}
{Schaye} J.,  {Dalla Vecchia} C.,  2008, \mnras, 383, 1210

\bibitem[\protect\citeauthoryear{{Schaye}, {Dalla Vecchia}, {Booth}, {Wiersma},
  {Theuns}, {Haas}, {Bertone}, {Duffy}, {McCarthy} \& {van de Voort}}{{Schaye}
  et~al.}{2010}]{Schaye}
{Schaye} J.,  {Dalla Vecchia} C.,  {Booth} C.~M.,  {Wiersma} R.~P.~C.,
  {Theuns} T.,  {Haas} M.~R.,  {Bertone} S.,  {Duffy} A.~R.,  {McCarthy} I.~G.,
     {van de Voort} F.,  2010, \mnras, 402, 1536

\bibitem[\protect\citeauthoryear{{Shaw}, {Weller}, {Ostriker} \& {Bode}}{{Shaw}
  et~al.}{2006}]{Shaw}
{Shaw} L.~D.,  {Weller} J.,  {Ostriker} J.~P.,    {Bode} P.,  2006, \apj, 646,
  815

\bibitem[\protect\citeauthoryear{{Sheth} \& {Tormen}}{{Sheth} \&
  {Tormen}}{2004}]{Sheth}
{Sheth} R.~K.,  {Tormen} G.,  2004, \mnras, 349, 1464

\bibitem[\protect\citeauthoryear{{Spergel}, {Bean}, {Dor{\'e}}, {Nolta},
  {Bennett}, {Dunkley}, {Hinshaw}, {Jarosik}, {Komatsu}, {Page} \& et
  al.}{{Spergel} et~al.}{2007}]{Spergel07}
{Spergel} D.~N.,  {Bean} R.,  {Dor{\'e}} O.,  {Nolta} M.~R.,  {Bennett} C.~L.,
  {Dunkley} J.,  {Hinshaw} G.,  {Jarosik} N.,  {Komatsu} E.,  {Page} L.,    et
  al. 2007, \apjs, 170, 377

\bibitem[\protect\citeauthoryear{{Spergel}, {Verde}, {Peiris}, {Komatsu},
  {Nolta}, {Bennett}, {Halpern}, {Hinshaw}, {Jarosik}, {Kogut} \& et
  al.}{{Spergel} et~al.}{2003}]{Spergel03}
{Spergel} D.~N.,  {Verde} L.,  {Peiris} H.~V.,  {Komatsu} E.,  {Nolta} M.~R.,
  {Bennett} C.~L.,  {Halpern} M.,  {Hinshaw} G.,  {Jarosik} N.,  {Kogut} A.,
  et al. 2003, \apjs, 148, 175

\bibitem[\protect\citeauthoryear{{Springel}}{{Springel}}{2005}]{Springel05}
{Springel} V.,  2005, \mnras, 364, 1105

\bibitem[\protect\citeauthoryear{{Springel}, {White} \& {Hernquist}}{{Springel}
  et~al.}{2001}]{Springel}
{Springel} V.,  {White} M.,    {Hernquist} L.,  2001, \apj, 549, 681

\bibitem[\protect\citeauthoryear{{Syer} \& {White}}{{Syer} \&
  {White}}{1998}]{Syer}
{Syer} D.,  {White} S.~D.~M.,  1998, \mnras, 293, 337

\bibitem[\protect\citeauthoryear{{van de Voort}, {Schaye}, {Booth}, {Haas} \&
  {Dalla Vecchia}}{{van de Voort} et~al.}{2011}]{Voort}
{van de Voort} F.,  {Schaye} J.,  {Booth} C.~M.,  {Haas} M.~R.,    {Dalla
  Vecchia} C.,  2011, \mnras, 414, 2458

\bibitem[\protect\citeauthoryear{{van den Bosch}}{{van den Bosch}}{2002}]{van}
{van den Bosch} F.~C.,  2002, \mnras, 331, 98

\bibitem[\protect\citeauthoryear{{van den Bosch}, {Jiang}, {Hearin},
  {Campbell}, {Watson} \& {Padmanabhan}}{{van den Bosch}
  et~al.}{2014}]{vandenBosch14}
{van den Bosch} F.~C.,  {Jiang} F.,  {Hearin} A.,  {Campbell} D.,  {Watson} D.,
     {Padmanabhan} N.,  2014, \mnras, 445, 1713

\bibitem[\protect\citeauthoryear{{Wang}, {Mo} \& {Jing}}{{Wang}
  et~al.}{2009}]{WangH}
{Wang} H.,  {Mo} H.~J.,    {Jing} Y.~P.,  2009, \mnras, 396, 2249

\bibitem[\protect\citeauthoryear{{Wang} \& {White}}{{Wang} \&
  {White}}{2009}]{Wang}
{Wang} J.,  {White} S.~D.~M.,  2009, \mnras, 396, 709

\bibitem[\protect\citeauthoryear{{Wechsler}, {Bullock}, {Primack}, {Kravtsov}
  \& {Dekel}}{{Wechsler} et~al.}{2002}]{Wechsler}
{Wechsler} R.~H.,  {Bullock} J.~S.,  {Primack} J.~R.,  {Kravtsov} A.~V.,
  {Dekel} A.,  2002, \apj, 568, 52

\bibitem[\protect\citeauthoryear{{Wiersma}, {Schaye} \& {Smith}}{{Wiersma}
  et~al.}{2009a}]{WiersmaA}
{Wiersma} R.~P.~C.,  {Schaye} J.,    {Smith} B.~D.,  2009a, \mnras, 393, 99

\bibitem[\protect\citeauthoryear{{Wiersma}, {Schaye}, {Theuns}, {Dalla Vecchia}
  \& {Tornatore}}{{Wiersma} et~al.}{2009b}]{WiersmaB}
{Wiersma} R.~P.~C.,  {Schaye} J.,  {Theuns} T.,  {Dalla Vecchia} C.,
  {Tornatore} L.,  2009b, \mnras, 399, 574

\bibitem[\protect\citeauthoryear{{Zhao}, {Jing}, {Mo} \& {B{\"o}rner}}{{Zhao}
  et~al.}{2009}]{Zhao09}
{Zhao} D.~H.,  {Jing} Y.~P.,  {Mo} H.~J.,    {B{\"o}rner} G.,  2009, \apj, 707,
  354

\bibitem[\protect\citeauthoryear{{Zhao}, {Mo}, {Jing} \& {B{\"o}rner}}{{Zhao}
  et~al.}{2003}]{Zhao03}
{Zhao} D.~H.,  {Mo} H.~J.,  {Jing} Y.~P.,    {B{\"o}rner} G.,  2003, \mnras,
  339, 12

\end{thebibliography}
\bibliographystyle{mn2e}

\appendix

\section{Simulations}\label{Sims_details}


The DMONLY simulations contain only dark matter particles, which provide us with a useful baseline model for testing the impact of baryonic physics on halo mass histories and mass profiles, when comparing it with the REF simulation. The REF simulation includes sub-grid recipes for star formation (\citealt{Schaye08}), radiative (metal-line) cooling and heating (\citealt{WiersmaA}), stellar evolution, mass loss from massive stars and chemical enrichment (\citealt{WiersmaB}), and a kinetic prescription for supernova feedback (\citealt{DallaVecchia}). In this work we use hydrodynamical simulations (along with DMONLY) to test the effect of baryons on halo mass histories and compute the gas accretion rate. We do not test other feedback schemes because it was shown by \citet{Voort} that the halo mass accretion is robust to variations in feedback.


\subsection{Construction of halo merger trees}\label{merger_tree}

The first step towards studying the mass assembly history of halos is to identify gravitationally bound structures and build halo merger trees. We begin by selecting the largest halo in each FoF group (\citealt{Davis}) (i.e. the main subhalo of FoF groups that is not embedded inside larger halos). Halo virial masses and radii are determined using a spherical overdensity routine within the SUBFIND algorithm (\citealt{Springel,Dolag}) centred on the main subhalo of FoF halos.  Therefore, we define halo masses as all matter within the radius $r_{200}$ for which the mean internal density is 200 times the critical density. Throughout this work, we study the accretion history of the largest halos in each FoF group. Subhalos, defined as bound structures that reside within the virial radius of the largest `host' halo, show distinct mass histories. The structures of subhalos are strongly affected by the potential of their host halos, as seen for example in the cessation of mass accretion on to subhalos residing in dense environments (see \citealt{WangH} or \citealt{Lacerna}). Consequently, the masses of subhalos do not follow the mass history of their host halos.

The merger trees of the largest halos are then built as follows. First, at each output redshift (snapshot), we select `resolved' halos that contain more than 300 dark matter particles. We refer to these resolved halos as `descendants'. We then link each descendant with a unique `progenitor' at the previous output redshift. This is nontrivial due to halo fragmentation: subhalos of a progenitor halo may have descendants that reside in more than one halo. The fragmentation can be spurious or due to a physical unbinding event. To correct this, we link the descendant to the progenitor that contains the majority of the descendant's 25 most bound dark matter particles. Therefore, the main progenitor of a given dark matter halo is found by tracing backwards in time to the most massive halo along the main branch of its merger tree. The different mass histories are calculated by following the merger trees of a given sample of halos. At each redshift the mass histories are computed by calculating the median mass value, determined by bootstrap resampling the halos, from the merger tree. Along with the median value, the $1\sigma$ confidence interval is recorded.

\begin{figure*} 
  \centering
  \subfloat{\includegraphics[angle=0,width=0.48\textwidth]{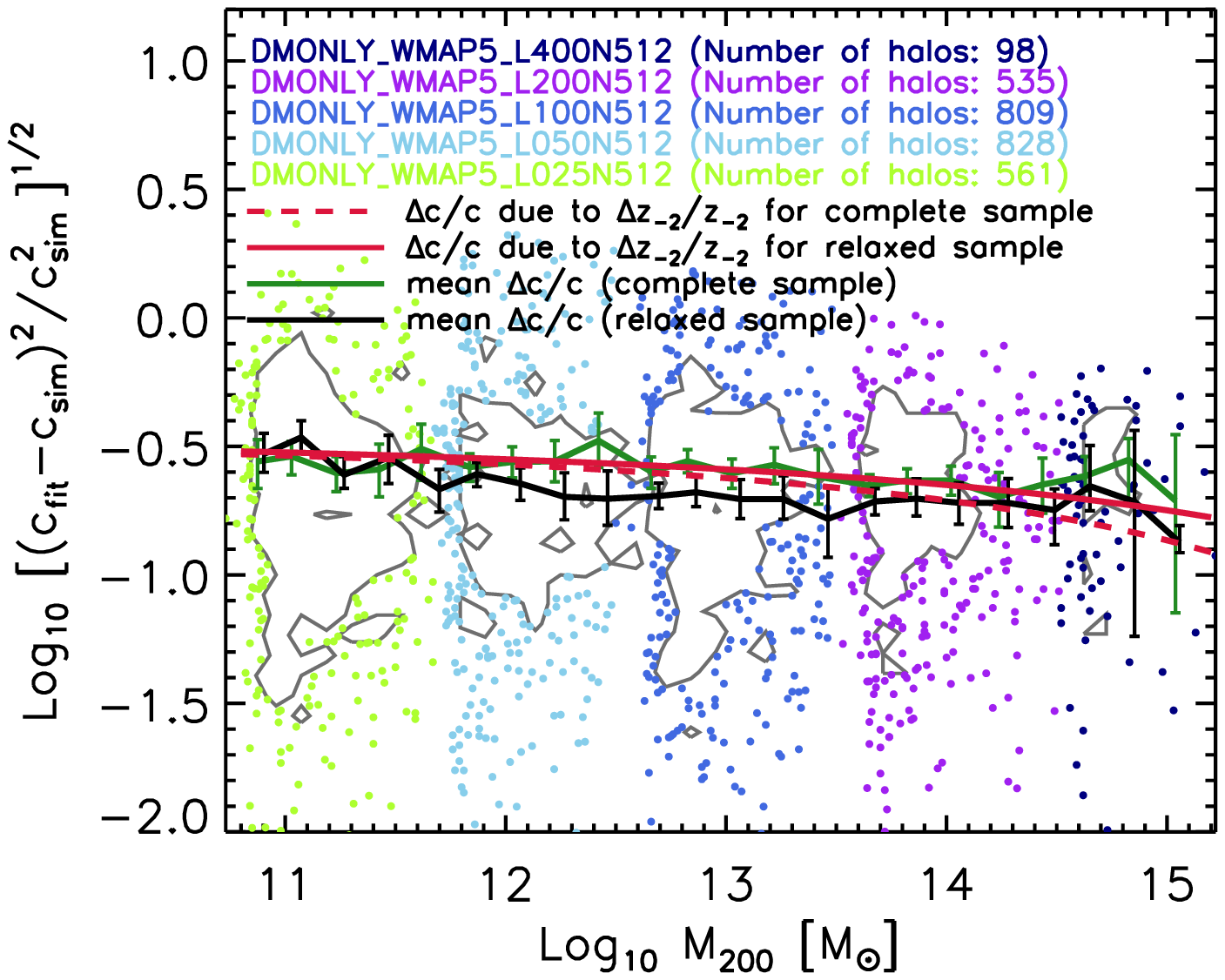}}
  \subfloat{\includegraphics[angle=0,width=0.48\textwidth]{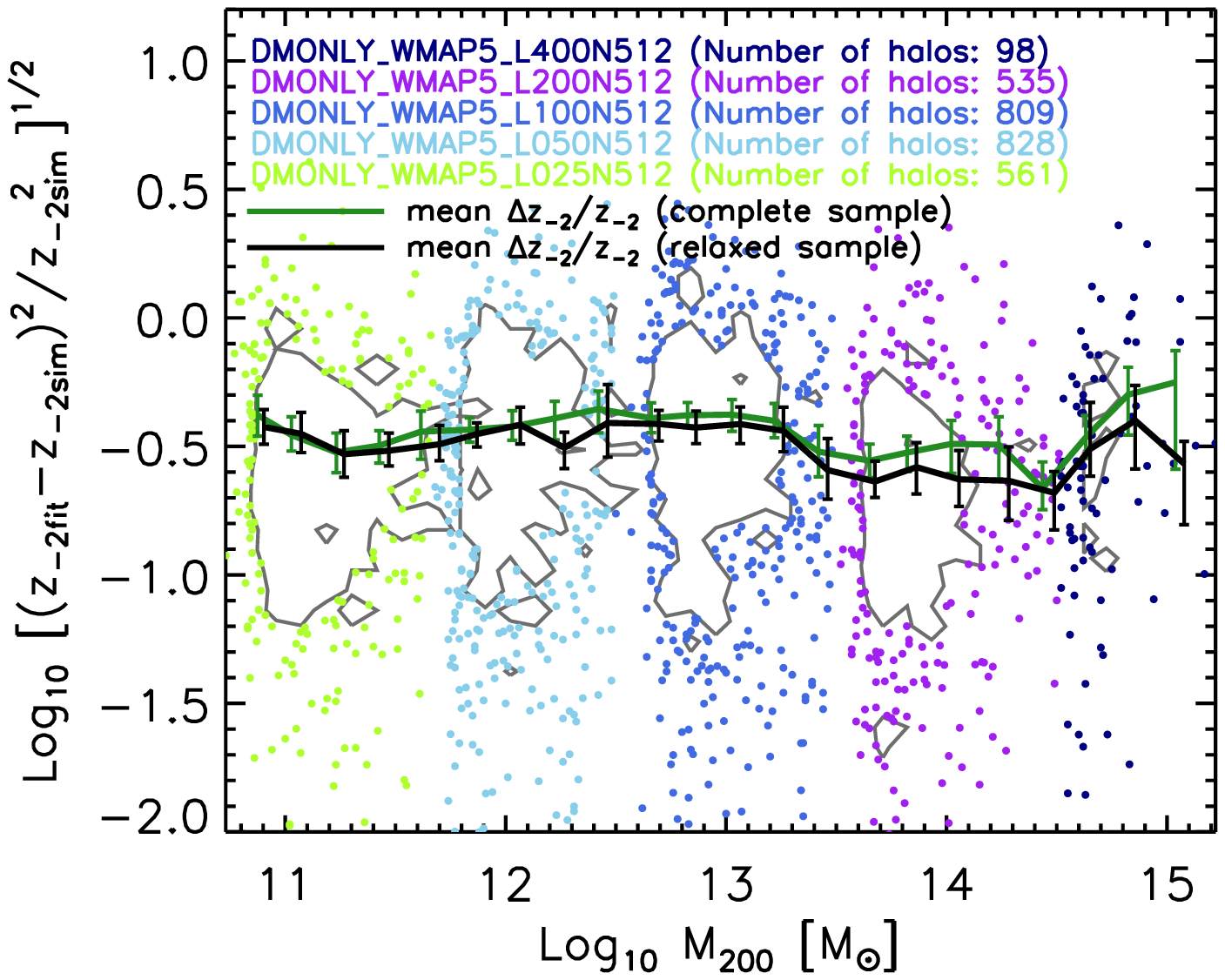}}
\caption{Scatter in the concentration$-$mass relation (left panel) and the formation redshift$-$mass relation (right panel). {\it{Left panel}}: Y-axis shows the difference between the concentration predicted by Duffy et al. (2008) (see eq.~\ref{c-m}) and the actual concentration from the simulation output. This difference is divided by the concentration from the simulation output and plotted against virial halo mass. The different colours of the points indicate that the concentration outputs were obtained from DMONLY$_{-}$WMAP5 simulations with different box sizes. The grey contours enclose $68\%$ of the distribution while the individual points show the remaining $32\%$. The black (green) solid line shows the mean relative scatter in the concentration-mass relation per halo mass bin of the relaxed (complete) sample. The red dashed line is an analytic estimate of the scatter obtained by propagating the scatter in the formation redshift$-$mass relation to the concentration$-$mass relation (eq.~\ref{scatter_eq}). {\it{Right panel}}: same as the left panel but the scatter is obtained from the difference in the formation redshift predicted from eq.~(\ref{zf-c}) and the simulation output. The black (green) solid line shows the median value in the scatter as a function of mass of the relaxed (complete) sample.}
\label{scatter_plot}
\end{figure*}

\section{Analysis of scatter}\label{Scatter_analysis}

\subsubsection{Scatter in formation times and concentrations}\label{Sec_scatter}

We now analyse the scatter in the formation time$-$mass relation and show it relates to the scatter in the concentration$-$mass relation. So far, we have shown that the formation time is related to halo concentration through the mean inner density ($\langle\rho\rangle(<r_{-2})$)$-$critical density ($\rho_{\rm{crit}}(z_{-2})$) relation plotted in Fig. \ref{density_relation}. Through first order error propagation, we look for the corresponding scatter in the formation time,

\begin{eqnarray}
900\rho_{\rm{crit}}(z_{-2}) &=& \langle\rho\rangle(<r_{-2}),\\
900\delta(\rho_{\rm{crit}}(z_{-2})) &=& \delta(\langle\rho\rangle(<r_{-2})),\\
\frac{900}{200}\delta[\Omega_{m}(1+z_{-2})^{3}+\Omega_{\Lambda}] &=& \delta\left(c^{3}\frac{Y(1)}{Y(c)}\right),\\\label{scatter_eq}
3\Omega_{m}(1+z_{-2})^{2}z_{-2}\left(\frac{\delta z_{-2}}{z_{-2}}\right) &=& \frac{\langle\rho\rangle(<r_{-2})}{900\rho_{0}}\left(\frac{\delta c}{c}\right)\\\nonumber
&& \times\left(3-\frac{c^{2}}{(1+c)^{2}}\frac{1}{Y(c)}\right),
\end{eqnarray} 

\noindent where we used $\delta(c^{3}/Y(c))=3c^{2}\delta c/Y(c)-c^{4}\delta c/[Y(c)^{2}(1+c)^{2}]$. Eq.~(\ref{scatter_eq}) relates the scatter in formation time ($|\delta z_{-2}|/z_{-2}$) to the scatter in the concentration ($|\delta c|/c$). 

The grey shaded areas in the panels in Fig. \ref{zformation} show the scatter in $z_{-2}$, while the panels in Fig. \ref{scatter_plot} show the scatter in the $c-M$ relation (left panel) and in the $z_{-2}-M$ relation (right panel). The grey contours in Fig.~\ref{scatter_plot} enclose $68\%$ of the distribution while the individual points show the remaining $32\%$. The black (green) solid line shows the mean scatter in the formation time per halo mass bin for the relaxed (complete) sample. The presence of unrelaxed halos does not have any significant effect on either the scatter in the formation time or the mass histories. The average scatter in formation time is $\langle|\delta z_{-2}|/z_{-2}\rangle=0.324$ ($\langle|\delta z_{-2}|/z_{-2}\rangle=0.356$) for the relaxed (complete) sample. 

The left panel of Fig. \ref{scatter_plot} shows that the average scatter in the concentration$-$mass relation is $\langle|\delta c|/c\rangle=0.257$ for the full sample and $\langle|\delta c|/c\rangle=0.218$ for the relaxed sample. In agreement with previous work (see e.g. \citealt{Neto}), the scatter in the concentration of the relaxed halo sample is lower than the scatter of the full sample. The extra scatter in the full sample is produced by the deviation of the density profiles from the NFW form for halos experiencing ongoing mergers and for artificially linked halos. 

Assuming $\langle|\delta z_{-2}|/z_{-2}\rangle=0.324$, $\langle|\delta c|/c\rangle$ can be obtained as a function of halo mass by applying eq. (\ref{scatter_eq}). This analytic estimate is plotted in the left panel for the relaxed sample (red dashed line). We find very good agreement between the scatter in concentration from eq.~(\ref{scatter_eq}), and the median value plotted in black for the relaxed sample, and in green for the complete sample. Therefore, we conclude that the scatter in formation time determines the scatter in the concentration. However, at higher masses and redshifts the fraction of relaxed halos decreases (e.g. \citealt{Ludlow12}) and it has been found that the concentration-mass relation of the complete halo sample exhibits a strong flattening and upturn (\citealt{Klypin,Prada}). As a result, at high masses and redshifts the scatter in the concentration will probably depend on other variables besides the scatter in formation time.

In the following subsection we find that the scatter in the accretion history determines the scatter in the formation time.

\begin{figure} 
\begin{center}
\includegraphics[angle=0,width=0.48\textwidth]{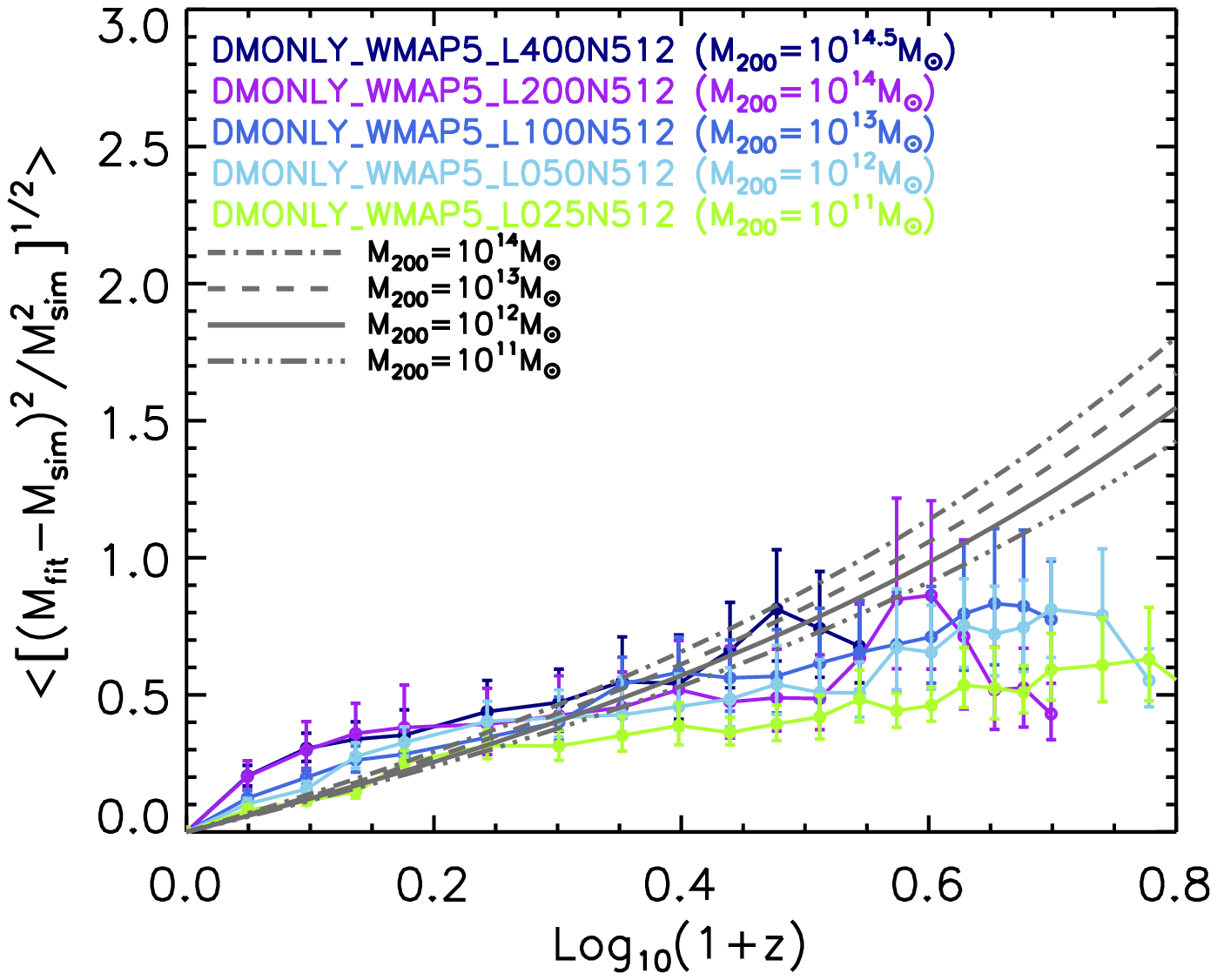}%
\caption{Mean scatter in the halo mass history against redshift for different halo masses $M_{200}$. The y-axis shows the mean value of the difference between the mass history ($M_{\rm{fit}}$) predicted by eqs.~(\ref{McBride}), (\ref{alpha2}) and (\ref{beta}), and the mass history ($M_{\rm{sim}}$) from the simulation output. The difference is divided by the $M_{\rm{sim}}$ from the simulation output. The different coloured lines correspond to the different DMONLY$_{-}$WMAP5 simulations as indicated in the legends. The grey lines show the analytic estimates of the mass history curves given by eqs.~(\ref{error}), (\ref{dalpha1}) and (\ref{dalpha2}).}
\label{MH_error}
\end{center}
\end{figure}

\subsubsection{Scatter in halo mass histories}

In this section we analyse the scatter found when computing mass histories from the simulation outputs. We analytically estimate the scatter in the mass history due to both the scatter in the concentration and formation times (estimated in Section \ref{Sec_scatter}). We then compare this to the scatter obtained from the simulation outputs.

To compute the scatter in the mass history we perform a first order error propagation in $M(z)$ (eq.~\ref{McBride}),

\begin{equation}\label{error}
\delta M(z)/M(z) = \delta\alpha\ln(1+z)+z\delta\beta,
\end{equation}

\noindent where $\delta\alpha$ ($\delta\beta$) is the scatter in $\alpha$ ($\beta$) due to the scatter in $c$ and $z_{-2}$. From eq.~(\ref{alpha2}) we first compute $\delta\alpha$ due to the scatter in $z_{-2}$,

\begin{eqnarray}\nonumber
\delta\alpha &=& \delta(-\beta z_{-2}/\ln(1+z_{-2}))\\\nonumber
&=& \frac{-z_{-2}\delta\beta-\beta\delta z_{-2}}{\ln(1+z_{-2})}-\beta z_{-2}\delta[\ln(1+z_{-2})]^{-1}\\\label{dalpha1}
&=& -\frac{z_{-2}\beta(1+z_{-2}\ln(1+z_{-2}))}{(1+z_{-2})\ln(1+z_{-2})}\left(\frac{\delta z_{-2}}{z_{-2}}\right),
\end{eqnarray}

\noindent where in the last line we used $\delta\beta = \frac{-\beta z_{-2}}{(1+z_{-2})}\left(\frac{\delta z_{-2}}{z_{-2}}\right)$, which follows from eq. (\ref{beta}). Similarly, we calculate the scatter in $\alpha$ due to the scatter in $c$,

\begin{eqnarray}\nonumber
\delta\alpha &=& \delta(\ln(Y(1)/Y(c))/\ln(1+z_{-2})\\\nonumber
&=& -\frac{\delta Y(c)}{Y(c)}\frac{1}{\ln(1+z_{-2})}\\\label{dalpha2}
&=& -\frac{c^{2}}{(1+c)^{2}Y(c)\ln(1+z_{-2})}\left(\frac{\delta c}{c}\right),\\\nonumber
\end{eqnarray}

\noindent where in the last step we used $\delta Y(c)=\frac{c^{2}}{(1+c)^{2}}\left(\frac{\delta c}{c}\right)$. Finally, we substitute $\delta\beta=\frac{-\beta z_{-2}}{(1+z_{-2})}\left(\frac{\delta z_{-2}}{z_{-2}}\right)$ and eqs.~(\ref{dalpha1}) and (\ref{dalpha2}) into eq.~(\ref{error}). From Section~\ref{Sec_scatter} we adopt the average scatter in $z_{-2}$ and $c$ for the complete sample, that is $\langle|\delta c|/c\rangle=0.25$ and $\langle\delta z_{-2}/z_{-2}\rangle=0.35$, and compute $\langle|\delta M(z)/M(z)|\rangle$ as a function of redshift. Fig. \ref{MH_error} shows the scatter measured from the simulation outputs (coloured lines) and from the analytic estimates (grey dashed lines). The different coloured lines correspond to the median values of the scatter from simulations with different box sizes. We calculate these by averaging the difference between the mass history predicted by eqs.~(\ref{McBride}), (\ref{alpha2}) and (\ref{beta}) (for a given halo with mass $M_{200}$ at $z=0$), and each $M(z)$ given by the merger trees from the complete halo sample. We find very good agreement indicating that the scatter in the concentration comes from the scatter in the formation time, which in turn comes from the scatter in the halo mass history.    

\section{Step-by-step guide to compute halo mass histories}\label{MAH_Description}

\subsubsection{Semi-analytical model}

This Appendix provides a step-by-step procedure that outlines how to calculate the halo mass histories using the numerical model presented in Section 4.4:

\begin{enumerate}

\item First assume a cosmology and choose a concentration-mass relation from the literature. For instance, the Duffy et al. (2008) relation, ${c=6.67(M_{0}/2\times 10^{12}h^{-1}\rm{M}_{\sun})^{-0.092}}$, is suitable for the WMAP5 cosmology, whereas \citet{Neto} is suitable for WMAP1:

\item Calculate the formation time,

\begin{equation}
z_{-2}=\left(\frac{200}{A_{\rm{cosmo}}}\frac{c(M_{0})^{3}Y(1)}{\Omega_{\rm{m}}Y(c(M_{0}))}-\frac{\Omega_{\Lambda}}{\Omega_{\rm{m}}}\right)^{1/3}-1.
\end{equation}

\noindent Note that the value of $A_{\rm{cosmo}}$ in the above equation is cosmology dependent. $A_{\rm{WMAP5}}=900$ is suitable for the WMAP5 cosmology. In this work we obtained $A_{\rm{WMAP1}} = 787$, $A_{\rm{WMAP3}} = 850$, $A_{\rm{WMAP9}} = 820$ and $A_{\rm{Planck}} = 798$.

\item Calculate the parameters $\alpha$ and $\beta$,

\begin{eqnarray}
\alpha &=& [\ln(Y(1)/Y(c))-\beta z_{-2}]/\ln(1+z_{-2}),\\
\beta &=& -3/(1+z_{-2}).
\end{eqnarray}

\item Finally, the mass history can be calculated as follows,

\begin{eqnarray}
M(z)=M_{0}(1+z)^{\alpha}e^{\beta z}.
\end{eqnarray}

\end{enumerate}

The above model is suitable for any cosmology (as long as the concentration-mass relation and the value of $A_{\rm{cosmo}}$ correspond to the desire cosmology) and is valid over the halo mass range for which the concentration-mass and the $z_{-2}-M_{0}$ relations, obtained from simulations, are valid (e.g. $10^{10}-10^{14}\rm{M}_{\sun}$ for Duffy et al. 2008).

\end{document}